\documentclass[prb,reprint,superscriptaddress,titlepage,floatfix,showkeys]{revtex4-2} 
\usepackage{graphicx}
\usepackage{gensymb}
\usepackage{nicefrac}
\usepackage{amsfonts}
\usepackage[version=3]{mhchem}
\usepackage{amssymb}
\usepackage{amsmath}
\usepackage{subfigure}
\usepackage{multirow}
\usepackage{tabularx}
\usepackage{array}
\usepackage{units}
\usepackage{braket}
\usepackage{bm,times}
\usepackage{booktabs}
\usepackage{enumitem}
\usepackage{mathtools}
\usepackage{epstopdf}
\usepackage[dvipsnames]{xcolor}
\usepackage{threeparttable}
\usepackage[colorlinks = true, linkcolor = blue, urlcolor  = blue, citecolor = blue, anchorcolor = blue]{hyperref}

\def\asi{{\it a}-Si} 
\def\csi{{\it c}-Si} 
 
\def\u2{$\langle u^2 \rangle$}
\def\ui{$u_i^2$}

\def\gmc{g\,cm$^{-3}$}
\def\bea{\begin{eqnarray}}
\def\eea{\end{eqnarray}}
\def\be{\begin{equation}}
\def\ee{\end{equation}}

\begin{document}

\title{
Ab initio studies of the impact of the Debye-Waller factor 
on the structural and dynamical properties of amorphous 
semiconductors: The case of a-Si.
} 

\author{Devilal Dahal}
\affiliation{
Department of Physics and Astronomy, University of Southern 
Mississippi, Hattiesburg, Mississippi 39406, USA}
\email{devilal.dahal@usm.edu}

\author{Raymond Atta-Fynn}
\affiliation{
Materials Science and Technology Division, Los Alamos National 
Laboratory, Los Alamos, New Mexico 87545, USA
}
\email{attafynn@lanl.gov}

\author{Stephen R. Elliott}
\affiliation{Physical and Theoretical Chemistry Laboratory,
University of Oxford, Oxford OX1 3QZ, United Kingdom}
\email{stephen.elliott@chem.ox.ac.uk}

\affiliation{Department of Chemistry, University of Cambridge,
Cambridge CB2 1EW, United Kingdom}
\email{sre1@cam.ac.uk}

\author{Parthapratim Biswas}
\email[Corresponding author:\,]{partha.biswas@usm.edu}
\affiliation{
Department of Physics and Astronomy, University of Southern 
Mississippi, Hattiesburg, Mississippi 39406, USA}

\begin{abstract}
This paper presents a first-principles study of the Debye-Waller 
factor and the Debye temperature for amorphous silicon ({\asi}) 
from lattice-dynamical calculations and direct molecular-dynamics 
simulations using density-functional theory (DFT).  The effects of 
temperature and structural disorder on the intensity 
of the diffraction maxima and the vibrational 
mean-square displacement (MSD) of Si atoms are studied 
in the harmonic approximation, with particular emphasis 
on the bond-length disorder, the presence of coordination 
defects, and microvoids in {\asi} networks. 
It has been observed that the MSDs associated with 
tetrahedrally-bonded Si atoms are considerably lower 
than their dangling-bond counterparts -- 
originating from isolated and vacancy-induced clustered 
defects -- and those on the surface of microvoids, leading 
to an asymmetric  non-gaussian tail in the distribution 
of atomic displacements.  An examination of the effect 
of anharmonicity on the MSD at high temperatures using 
direct {\it ab initio} molecular-dynamics simulations 
(without the harmonic 
approximation) suggests that the vibrational motion in {\asi} is practically 
unaffected by anharmonic effects at temperatures below 
400~K, as far as the present DFT calculations are concerned.  
The Debye temperature of {\asi} is found to be in the 
range of 488--541~K from specific-heat and MSD 
calculations using first-principles lattice-dynamical 
calculations in the harmonic approximation, which 
matches closely with the experimental value of 487--528~K 
obtained from specific-heat measurements of {\asi} at 
low temperatures. 
\end{abstract}

\maketitle

\section{Introduction} 
The influence of thermal vibrations on the intensity
of diffraction maxima in crystals has been extensively
studied in the literature~\cite{Debye:1913,Faxen:1923,Waller:1923,
Ott:1935,Born:1942}.
Following Debye's original work~\cite{Debye:1913}, where
vibrations in crystals were assumed to be
independent of the position of atoms in a lattice,
more accurate calculations by Fax{\'e}n~\cite{Faxen:1923}
and Waller~\cite{Waller:1923} showed that thermal
vibrations reduce the intensity of the diffraction maxima
but do not affect the sharpness of the diffraction lines.
The missing intensity from the spectra was found
to be present in the general background, which is
known as the temperature diffuse scattering (TDS).
The intensity reduction factor,
$e^{-2M}$, is known as the Debye-Waller factor~\cite{DW}, where
$M = 8\pi^2 \langle {u^2}\rangle \sin^2 \theta/\lambda^2$
for elemental solids.
The mean-square displacement (MSD) of atoms, {\u2}, along a
direction ${\mathbf Q}$, is perpendicular to the reflecting
plane and  $\theta$ is the glancing angle of incidence 
with the plane.
The vector ${\mathbf Q}$ is given by (${\mathbf k}-{\mathbf k_0}$),
where ${\mathbf k}$ and ${\mathbf k_0}$ are the wavevectors for
the scattered and incident beams, respectively. In Bragg
geometry, the scattering angle $2\theta$ is given by the angle
between $\mathbf k$ and $\mathbf k_0$, and
$|{\mathbf k}|$ = $|{\mathbf k_0}|$ for elastic scattering
with $|{\mathbf Q}| = 2|{\mathbf k}|\sin\theta =
4\pi\sin\theta/\lambda$, where $\lambda$ is the wavelength
of the scattering radiation.

The relationship between the Debye-Waller factor and 
finite-temperature atomic vibrations in solids was 
extensively studied in the last century~\cite{Debye:1913,Faxen:1923,
Waller:1923,Ott:1935,Born:1942,Shukla:1989,Shukla:1992,
Reid:1980,Flensburg:1999,Rignanese:1996,Lee:1995}.
However, the great majority of these studies are primarily focused
on elemental and molecular {\em crystals} using a variety of 
theoretical methods. Among these methods, the so-called 
statistical-moment approach~\cite{Hung:2014}, Green function
techniques~\cite{Shukla:1989,Shukla:1992},
and lattice-dynamical calculations, employing valence
force fields, shell models, and adiabatic
bond-charge models~\cite{Reid:1980,Flensburg:1999}, are
particularly noteworthy. In recent years, the
calculation of the Debye-Waller factor of crystalline materials
has been addressed by using total-energy and forces
from density-functional theory (DFT)~\cite{Vila:2007,Rignanese:1996,Lee:1995}.
By contrast, there exist only few studies that address 
the effect of the Debye-Waller factor in the amorphous 
state, e.g. As, SiO$_2$ and InP~\cite{Lottici:1987,Taraskin:1999, Schnohr:2009}.
Unlike elemental crystals, the MSD of an atom in an amorphous 
solid varies from site to site and depends on the local topology 
of the network.  As a result, the MSD of amorphous solids 
can significantly depend on the structural quality 
of the models employed in the calculations. It is 
therefore absolutely necessary to compute the 
Debye-Waller factor using high-quality structural 
models of amorphous solids obtained from {\it ab initio} 
density-functional calculations. 
A number of material properties, for example, the 
first sharp diffraction peak (FSDP) and the principal 
peak of the diffraction intensity spectrum can be 
affected by the presence of thermal vibrations. 
For amorphous materials, these changes are particularly 
important as the diffraction data or the structure 
factor plays an important role in characterizing the 
structure of the materials.

The absence of translational symmetry in amorphous solids
means that the normal modes and frequencies of vibrations
have to be determined in the coordinate space by
computing the environment-dependent atomic force constants
without the benefit of any symmetry considerations. 
It is therefore necessary to calculate the 
Debye-Waller factor of amorphous solids in real space. 
Since the MSD of an atom at high temperatures can 
be shown to be proportional to the sum of the inverse 
square of the normal-mode frequencies in the harmonic 
approximation (see section IIB), it is also necessary 
to employ a reasonably large model of amorphous solids 
in an effort to include the contribution of atomic 
displacements from low-frequency normal modes of the 
system.  This requirement can pose a major computational 
impediment to accurate {\it ab initio} calculations 
of the MSD of atoms in real space for large models 
using an extended set of basis functions~\cite{DZ}. 
In this paper, we undertake such a task and study 
the effect of local disordering and atomic 
coordination on the MSD of atoms in amorphous 
silicon ({\asi}). The effect of the Debye-Waller 
factor on the intensity of diffraction peaks is 
studied, with emphasis on the first sharp diffraction 
peak (FSDP) and the principal peak of {\asi}. 
The variation of the MSD of atoms with temperature 
in the presence of coordination defects and short-range 
and medium-range ordering in {\asi} networks is 
discussed.  We also examine the effect of the 
temperature-induced anharmonicity on the vibrational 
dynamics of Si atoms in {\asi} at high temperatures 
and the possible role of low-frequency vibrations on the 
molar specific heat of {\asi} at low temperatures. 
Specifically, the Debye temperature of {\asi} is 
computed using the harmonic approximation and 
the results are compared with those obtained 
from experiments. This is particularly challenging 
for non-crystalline solids using {\it ab initio} 
density-functional calculations in real space.  
The use of the Debye approximation entails that 
one must employ sufficiently large models of 
{\asi} to compute the low-frequency vibrational 
modes which play a major role in the contribution 
of the specific heat at low temperatures. 

The rest of the paper is arranged as follows.  
Section~II is devoted to the development of 
model {\asi} networks and incorporation of 
defects and extended-range inhomogeneities, which 
are followed by quantum-mechanical lattice-dynamical calculations 
of the MSD of atoms in real space in the harmonic 
approximation, and direct {\it ab initio} molecular 
dynamics (AIMD) simulations without the use of 
the harmonic approximation. Section~III 
presents results and discussion, where the 
role of short-range ordering and 
coordination defects on the MSD of atoms 
is examined at low and high temperatures. 
The Debye temperature and the molar specific heat 
of {\asi} are calculated and compared with 
the corresponding experimental values from 
the literature.  A discussion of 
the possible role of the low-frequency vibrations 
of {\asi} and their effects on the molar specific 
heat is presented and the effect of 
anharmonicity on the MSD of atoms at high 
temperatures from direct AIMD calculations 
is discussed in this section. This is followed 
by the conclusions in section~IV. 

\section{Method and Models}
In this study, we have used a set of high-quality {\asi} models 
obtained from a combination of classical and AIMD 
simulations. The term `high-quality' 
here refers to the fact that structural, electronic, and vibrational 
properties of the models are in good agreement with 
experimental data and that the models are free from any coordination 
defects for all model sizes. 
Below, we give a brief description of the simulation 
method for model construction, which is followed by lattice-dynamical 
calculations of the MSD of atoms in the harmonic approximation. 
The calculation of the MSD from direct AIMD simulations that 
take into account the volume expansion of {\asi} and the 
anharmonicity in atomic forces at high temperatures is also 
discussed in this section. 

\subsection{Generation of {\asi} models} 

The simulation method employed here consists of two 
steps. In the first step, several 500-atom and two 2000-atom 
defect-free configurations, confined in a cubic simulation cell 
with the experimental mass density of {\asi} of 
2.28~{\gmc}~\cite{Custer:1994}, were produced using the 
modified Stillinger-Weber (SW) potential~\cite{Vink:2001-JNCP} 
via classical MD simulations. The simulations were 
performed in canonical ensembles with a time step 
of 1~fs, and the temperature of the system was maintained by a chain 
of Nos{\'e}-Hoover thermostats~\cite{Nose:1984,Hoover:1985}.  The second step 
involved thermalization of the resulting classical models at 300~K 
using AIMD simulations.  The AIMD runs were conducted for 5~ps 
using the density-functional code {\sc Siesta}~\cite{Siesta:2002} 
by employing a set of double-zeta (DZ) basis functions~\cite{DZ}. 
The use of the DZ basis functions is necessary for the 
calculation of the mean-square displacement (MSD) and the 
low-frequency vibrational modes of atoms, which were found 
to be dependent on the choice of 
the basis functions.  The total energy and forces were computed by solving the 
Kohn-Sham equation in the self-consistent field approximation, 
and the exchange-correlation contribution to the total energy 
was obtained by using the generalized gradient approximation 
(GGA). The resulting configurations from the AIMD runs were 
further relaxed geometrically via the limited-memory 
Broyden-Fletcher-Goldfarb-Shanno algorithm~\cite{BFGS:1987} to obtain a set of final 
structures for studying the Debye-Waller factor in {\asi}. 
A detailed discussion of the method employed here and 
the validation of the resulting models can be found 
in Ref.~\onlinecite{Ray:2018}. 
For the calculation of the MSD of the (approximate) 
tetrahedral sites in {\asi}, we have used two 
independent 500-atom models of {\asi} with 
no coordination defects in the networks. 

In order to examine the effect of coordination defects and 
extended-range inhomogeneities on the Debye-Waller factor 
of {\asi}, we have also studied a few defective models 
with dangling bonds and voids. 
Specifically, we have used five independent 
500-atom configurations with 2--3 at.~\% of dangling bonds 
and two additional independent 500-atom configurations 
with no coordination defects but a pair of spherical 
voids of radius 4 {\AA} to generate robust MSD 
statistics.  The dangling bonds (DB) studied in this 
work are of two types. The first type of DBs are 
vacancy-induced, which can be produced by removing a tetrahedrally-bonded Si atom 
from a 100\% defect-free {\asi} network. The removal of 
a tetrahedrally-bonded atom creates four neighboring dangling bonds (DB) 
at the vacancy site. Several such quartets of DBs were 
introduced in the network and the resulting network 
was annealed at 300~K (and 600~K) for a time 
period of 5 ps, followed by {\it ab initio} total-energy 
relaxation of the networks using {\sc Siesta}. Care 
was taken to ensure that at least 2--3 at.~\% of the 
total DBs persist in the final relaxed configurations. 
The second type of DBs studied here are sparsely distributed in the 
network, with no neighboring DBs in the vicinity of 10 {\AA}. 
Since direct MD simulations cannot produce these isolated DBs 
in a controlled manner, we have employed an accelerated 
metadynamics simulation method to generate these models. 
Metadynamics simulations of {\asi}~\cite{Meta2016} can 
produce configurations with a given concentration of 
$n$-fold-coordinated Si atoms, with $n$=2--5, for 
generating a sparsely distributed DBs in the networks.  
A description of the method in the context of 
simulating {\asi} models via metadynamics simulations 
was discussed by three of us in Ref.~\onlinecite{Meta2016}. 
The configurations obtained from metadynamics simulations 
were thermalized at 300~K (and 600~K) 
for 5 ps, followed by {\it ab initio} total-energy relaxation. 
Likewise, two independent 100\% defect-free {\asi} models 
were used to produce models with a pair of nanometer-size 
voids by thermalizing and relaxing the configurations.  
The final configurations were used to study the effect 
of voids on the MSD of Si atoms.

\subsection{Debye-Waller factor from lattice-dynamical calculations} 

To study the temperature dependence of the MSD 
of atoms in a vibrating solid, we have taken 
two distinct approaches.  The first approach involves lattice-dynamical 
calculations in the harmonic approximation, 
whereas the second approach relies on direct AIMD 
simulations in canonical and microcanonical ensembles. 
The latter enables us to include some aspects of 
temperature-induced anharmonic effects that can appear 
at high temperatures. Assuming that $u_{i\alpha}(t)$ 
is the displacement of atom $i$ along the $\alpha$ 
direction, where $\alpha = (x, y, z)$, the potential 
energy of the vibrating system can be written as 
a Taylor expansion about the equilibrium positions 
of the atoms.  Neglecting the cubic and higher-order 
terms for small atomic displacements and noting 
that the linear terms vanish at the equilibrium 
position, the equations of motion can be written as 
\be
m_i {\ddot{u}}_{i\alpha}(t) = 
- \sum_{j\beta} K_{i\alpha, j\beta} \: u_{j \beta}. 
\label{EQ1}
\ee
The coefficient $K_{i\alpha,j\beta}$ is an element 
of the force-constant matrix and it denotes the 
magnitude of the force acting on the $i$-th atom 
along the $\alpha$ direction when the $j$-th atom 
is displaced by a unit distance along the $\beta$ 
direction. Substituting $v_i = \sqrt{m_i}u_i$, 
and assuming the solution of (\ref{EQ1}) to 
be simple harmonic, $v(t) = v_0 \exp(-\imath\omega t)$, 
the system of linear equations in (\ref{EQ1}) 
can be expressed in a matrix form
\be 
{\mathbf D} - \omega^2 {\mathbf I} = \mathbf{0}. 
\label{EQ2}
\ee
Here, $\mathbf D$ is a real symmetric $3N \times 3N$ matrix, 
which is known as the mass-adjusted force-constant matrix 
or the dynamical matrix, $D_{i\alpha,j\beta} 
= K_{i\alpha,j\beta}/\sqrt{m_i m_j}$, and ${\mathbf I}$ is 
the identity matrix. The eigenvalues and the normalized 
eigenvectors of ${\mathbf D}$ give the squared frequencies 
($\omega^2$) and the polarization vectors ($\hat{\mathbf e}$) 
of the atoms for the normal modes, respectively. For a 
system in stable mechanical equilibrium, all the eigenvalues of $\mathbf D$ 
are positive. The atomic displacement at site $i$ is 
obtained from a linear combination of the normal modes 
\begin{flalign}
{\bf u}_{i\alpha}(t) = 
\frac{1}{\sqrt{m_i}} \sum_n A_0(n)\, 
\hat{\bf e}_{i\alpha}(n)\, e^{-\imath\omega_n t} \: \: (\alpha = x, y, z). 
\label{EQ3} 
\end{flalign}
In Eq.~(\ref{EQ3}), $A_0(n)$ is the vibrational amplitude 
for the $n$-th normal mode, which may include a phase 
factor, and $\mathbf{\hat{e}}_{i\alpha}(n)$ are the three 
polarization vectors of atom $i$, for $\alpha = (x, y, z)$, 
associated with the mode $n$. The value of $A_0(n)$ is 
indeterminate from Eq.~(\ref{EQ2}), but it can be obtained 
by calculating the average kinetic/potential energy of 
the system in thermal equilibrium. Assuming $\mathcal{T}$ 
is the average kinetic energy, we have  
\be 
\mathcal{T} = \left\langle \sum_{i,\alpha} \frac{1}{2}m_i \dot{u}^2_{i,\alpha} \right \rangle = \left \langle \sum_{n} \frac{1}{2} \omega^2_n A^2_0(n) \right \rangle  
= \sum_n \frac{1}{2} \left \langle E_n \right \rangle
\label{EQ4}
\ee 
where the last step in Eq.~(\ref{EQ4}) follows from the 
non-interacting nature of the normal modes obtained in the 
harmonic approximation. 
A similar calculation shows that the average potential energy, 
$\mathcal{V}$, of the system is also given by Eq.~(\ref{EQ4}). 
Since the normal modes can be treated as a set of independent 
harmonic oscillators, the average energy 
$\langle E_n \rangle$ for each mode in thermal equilibrium at 
temperature $T$ is given by the quantum-mechanical 
expression 
\be 
\langle E_n \rangle = \hbar\omega_n\left[\frac{1}{2} + 
\frac{1}{\exp{(\hbar\omega_n/k_BT)}-1}\right]
\label{EQ5} 
\ee
and the MSD follows from Eqs.~(\ref{EQ3})--(\ref{EQ5})
\begin{flalign} 
u^2_{i\alpha}(T)
& = \sum_n \frac{\langle E_n \rangle}{m_i\omega_n^2}\, 
|\hat{\mathbf e}_{i\alpha}(n)|^2 \label{EQ6} \\
& = 
\sum_n \frac{\hbar\, |\hat{\mathbf e}_{i\alpha}(n)|^2}{m_i\omega_n} 
\left[\frac{1}{2} + \frac{1}{\exp(\hbar\omega_n/k_BT) - 1} \right]. 
\label{EQ7}
\end{flalign} 
The MSD, $\langle u^2 \rangle$, can be readily obtained from averaging 
over all atoms and coordinate directions and the Debye-Waller factor 
follows from 
\be
M = \frac{8\pi^2\sin^2\theta}{\lambda^2} \langle u^2 \rangle, 
\label{EQ8}
\ee
where $\lambda$ and $2\theta$ are the wavelength of the 
scattering radiation and the angle of scattering, 
respectively.

At high temperatures, when $\hbar\omega_n/k_BT \ll 1$, 
one obtains the classical expression of the MSD 
by substituting $\langle E_n \rangle \approx k_BT$ in Eq.~(\ref{EQ6}). 
The first term in Eq.~(\ref{EQ7}) gives the 
contribution to the MSD from the zero-point motion (ZPM)  
of atoms, which leads to weak inelastic scattering, 
even at absolute zero temperature~\cite{Kittel}. The 
effect of thermal vibrations of atoms is reflected 
in the second term. The computation of the MSD in the 
lattice-dynamical approach can now be summarized as 
follows: (i) Thermalize the models at each temperature 
of interest for 5 ps, followed by {\it ab initio} 
total-energy optimization to prepare the system for 
the calculation of the ${\mathbf D}$ matrices; (ii) Construct 
the ${\mathbf D}$ matrices numerically in the harmonic 
approximation using {\it ab initio} forces, by displacing 
each atom, say  by 0.005 {\AA}, along the six coordinate 
directions ($\pm x$, $\pm y$, 
and $\pm z$); (iii) Diagonalize ${\mathbf D}$ 
to obtain the squared frequencies and the polarization 
vectors of the atoms for each mode, and calculate the 
MSD from Eq.~(\ref{EQ7}).  To produce good statistics, 
the results in section~IIIA were obtained from defect-free 
networks by averaging over two independent configurations, 
whereas those in section~IIIB are obtained from five to 
ten independent configurations with 2--3\% dangling 
bonds in the network.

\subsection{Debye-Waller factor from direct AIMD simulations}

The lattice-dynamical approach presented earlier does 
not include any anharmonic effects and it is suitable for 
temperatures well below the Debye temperature of 
solids. The effects of anharmonicity on the vibrational 
dynamics of atoms in {\asi} can be studied via direct 
AIMD simulations in canonical and microcanonical ensembles. 
In this approach, the system is first equilibrated 
at a given temperature in canonical ensembles so that 
the vibrational dynamics of the atoms are reflective 
of any temperature-induced structural changes that may 
take place in the system. Once the system is in 
equilibrium at a given temperature in canonical ensembles, 
it is then allowed to evolve in microcanonical ensembles. 
The use of microcanonical ensembles maintains the 
hamiltonian structure of the dynamics and conserves the 
total energy of the system in the absence of any 
thermostatting mechanism. Since the use of constant-volume 
NVE ensembles can partly restrict the system to include 
the effect of anharmonicity due to thermal expansion, 
the volume of the system before the NVE runs was adjusted 
on an ad hoc basis for each temperature as \be 
V(T) = V(T_0) [1+ \gamma (T - T_0)]^3,
\label{EQ9}
\ee  
where $\gamma$ is the coefficient of linear expansion 
of {\asi}, $V(T)$ is the volume of the system at 
temperature $T$, and $T_0$ = 300~K. For annealed samples of {\asi}, 
the experimental value of $\gamma$ is of the order 
of $4\times 10^{-6}$ K$^{-1}$\cite{Takimoto}. 
This value leads to a change of volume of about 0.36\% at 600~K from 
the original volume at $T_0$ = 300~K. Thus, it is 
unlikely that a change of temperature from 300~K 
to 600~K would induce a notable change of MSD values 
due to the volume expansion of the solid. 

The MSDs of the atoms were calculated from 
equilibrium microcanonical trajectories, and 
averaging the results over time and independent configurations during 
microcanonical runs. The MSD of atom $i$ at temperature 
$T$ can be written as
\be 
\langle u^2_i(T) \rangle = 
\langle 
\langle 
(r_i(t, T) - r_i^0(T))^2 
\rangle_{t}
\rangle_{\text{config}},
\label{EQ10}
\ee
where $r_i(t)$ is the position of atom $i$ at time $t$ and 
the symbol $\langle\ldots\rangle_{X}$ denotes averaging 
with respect to a variable $X$. Since the approach 
involves conducting long AIMD simulations in canonical 
and microcanonical ensembles, using double-zeta (DZ) basis 
functions for several independent configurations and 
temperatures, it becomes computationally prohibitive for 
large system sizes. We shall see later in section IIID that 
the use of extended or DZ basis functions is of 
paramount importance to accurately calculate the MSD 
and the molar specific heat of {\asi}. We have therefore 
restricted our simulations to three 216-atom models of {\asi} 
for the calculation of the MSD. The canonical and 
microcanonical runs were conducted for a time period of 
10 ps each, with a time step of 
1 fs, and the atomic trajectories were collected by 
evolving the system for an additional 10-ps microcanonical 
run beyond equilibration. The MSDs of the atoms were then 
calculated from Eq.~(\ref{EQ10}) for three independent 
configurations at several temperatures in the range from 
300~K to 600~K.

\section{Results and Discussion} 

\subsection{Temperature dependence of the Debye-Waller 
factor in {\asi}}

\begin{figure}[t!]
\centering
\includegraphics[width=0.8\columnwidth]{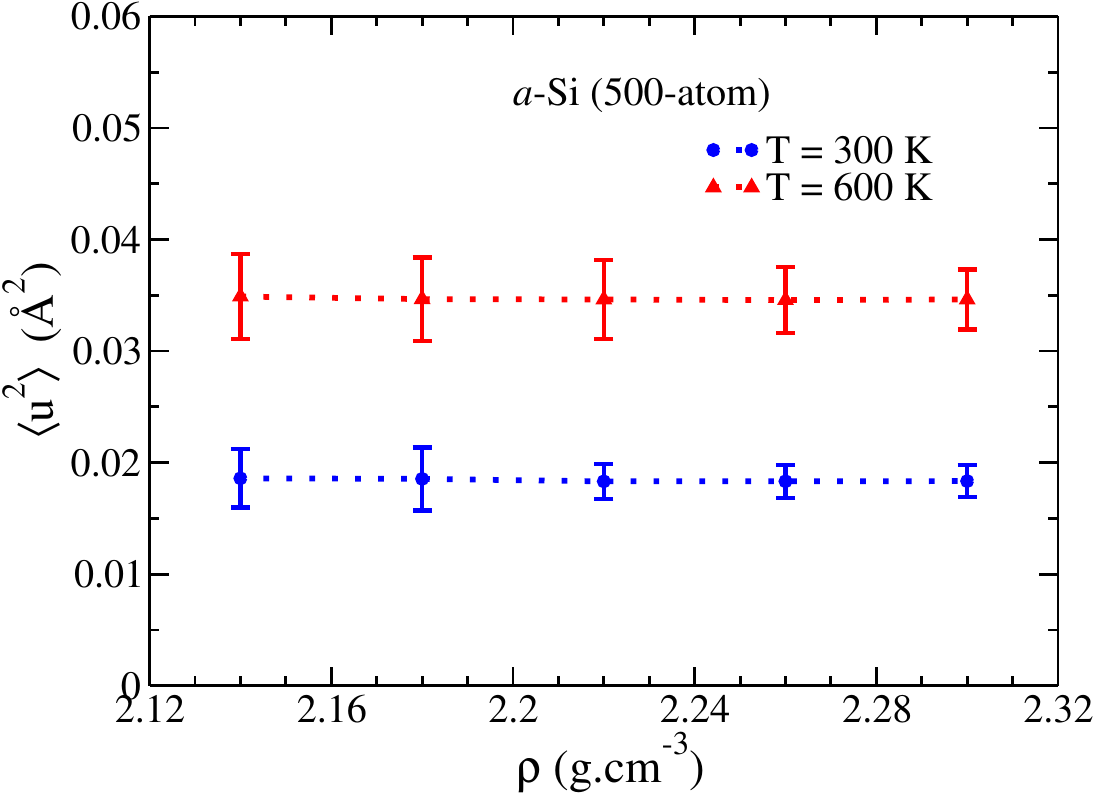}
\caption{
The MSD, {\u2}, of Si atoms in {\asi} with its density in the 
harmonic approximation from lattice-dynamical calculations. 
The MSD values can be seen to be practically 
independent of the density at 300~K (blue) and 
600~K (red).
}
\label{F1}
\end{figure}

\begin{figure}[t!]
\centering
\includegraphics[width=0.8\columnwidth]{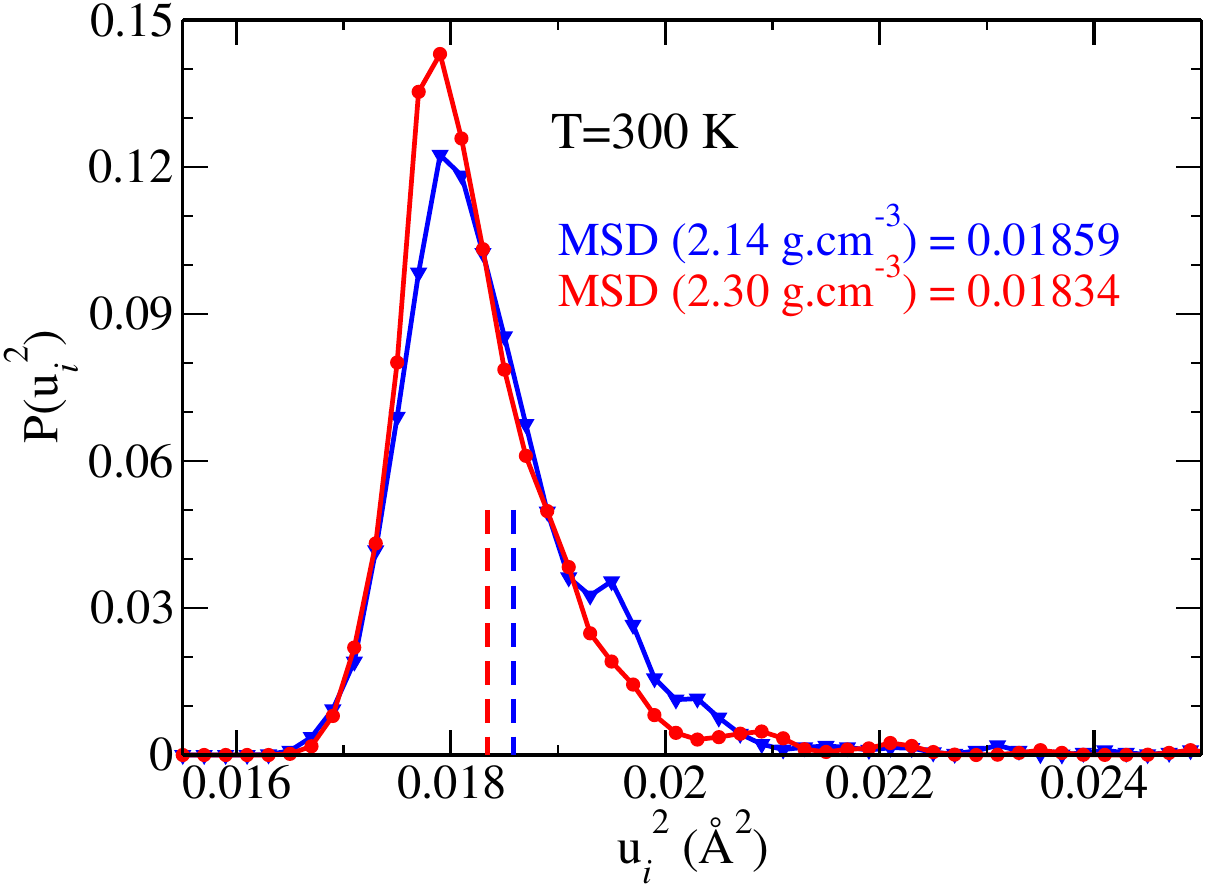}
\caption{
A comparison of the distributions of {\ui} of atoms 
in a low-density model (blue) and a high-density 
model (red) of {\asi} at 300~K obtained from two 
independent models with no coordination defects.  
The MSD values for the distributions are indicated 
as dashed vertical lines.  
}
\label{F2}
\end{figure}

In discussing our results, we begin with the variation of 
the MSD of atoms with the density of model {\asi} networks.  
Since the density of {\asi} samples can depend 
on preparation methods and experimental conditions, which 
may affect the local structure of {\asi} and hence local 
atomic force constants, it is important to examine whether the 
vibrational motion of Si atoms is sensitive to the density 
of {\asi} in the temperature range of 300--600~K. 
Figure \ref{F1} shows the variation, or the lack thereof, 
of the MSD, {\u2}, of atoms with the density of {\asi} 
at 300~K and 600~K. The data correspond to the average 
values obtained from two independent 500-atom model 
configurations.  The results suggest that the MSD is 
almost independent of the density of {\asi} within the 
range of 2.14--2.30 g\,cm$^{-3}$.  
This observation is not surprising, noting that the 
models studied in this work have no coordination 
defects and that a small variation of the density -- obtained 
via a homogeneous scale transformation 
of atomic distances, followed by thermalization 
at 300~K (and 600~K) for 5 ps and total-energy 
optimization -- from 2.3 to 2.14 g\,cm$^{-3}$ affects atomic distances 
by a linear scale factor of $s$ = 1.024.  Assuming that the 
system behaves as a harmonic solid, it can be shown 
that the MSD remains practically unchanged in the 
long-wavelength limit under a scale transformation 
from $\mathbf r \to s\mathbf r$ in the nearest-neighbor 
approximation between atoms~\cite{scale}. However, a small change 
of the MSD may result from the high-frequency modes 
and the deviation from the nearest-neighbor approximation 
of the atomic force constants. This 
is apparent from the results shown in Fig.~\ref{F2}, 
where a small shift of the MSD is found to lie 
well within one standard deviation of the distribution. 
It may be noted that the results presented in 
Fig.~\ref{F1} for 300~K have been explicitly 
verified for the two terminal densities of 
2.14 and 2.3 g\,cm$^{-3}$ by generating two 
independent models from random configurations 
and computing the MSD of the atoms for the 
resulting {\asi} models.

\begin{figure}[t!]
\includegraphics[width=0.8\columnwidth]{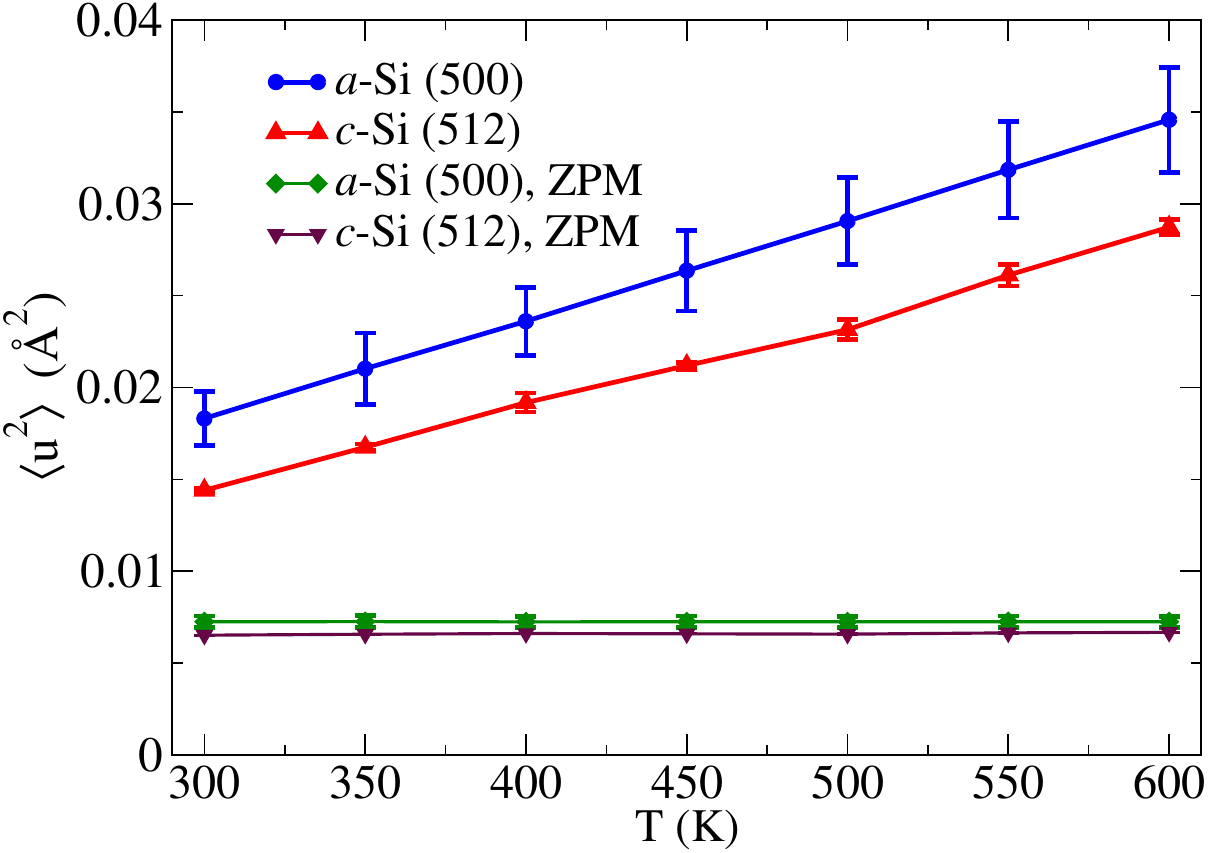}
\caption{
The variation of the MSD, {\u2}, with temperature 
in the harmonic approximation for {\asi} and 
diamond-structure {\csi}. The contribution to 
the MSD from the zero-point motion (ZPM) of the 
atoms is explicitly shown as horizontal lines. 
}
\label{F3}
\end{figure}

In the lattice-dynamical approach, the dependence of 
the MSD on temperature is generally studied in the 
harmonic approximation.  Elementary calculations in 
section~IIB show that, for small oscillations in the 
classical limit, the contribution to the MSD from 
thermal vibrations is directly proportional to the 
temperature of the system. 
Figure \ref{F3} shows the variation of the MSD, {\u2}, 
with temperature for model {\asi} networks 
with a density of 2.28 {\gmc}. The contributions 
to {\u2} from the zero-point motion (ZPM) and thermal 
vibrations of the atoms are shown separately in the 
figure. The MSD values due to thermal vibrations can 
be seen to increase linearly with temperature in 
Fig.~\ref{F3}, an observation which is consistent 
with the theoretical results obtained from the 
harmonic approximation in the temperature range 
of 300--600~K. The linear behavior also signifies 
that the normal-mode frequencies themselves are practically 
independent of the temperature in the range of 
300--600~K. For comparison, the MSD values for 
diamond-structure {\csi} are also presented in Fig.~\ref{F3}.  
A somewhat higher value of {\u2} in {\asi} compared 
to {\csi} can be attributed to the disorder associated 
with the local tetrahedral environment of {\asi}. 
In {\asi}, the atoms are bonded to four neighboring 
atoms in an approximate tetrahedral arrangement, 
which is characterized by the disorder in the 
bond-angle and bond-length distributions.  The 
presence of disorder reduces the strength of 
Si--Si bonds in the amorphous phase. This is in contrast to 
{\csi}, where Si atoms are strongly bonded to each 
other in an ideal tetrahedral arrangement.  This strong 
and compact ideal tetrahedral bonding results in 
slightly stiffened atomic force constants (and relatively 
high values for normal-mode frequencies in the harmonic 
approximation) compared to its amorphous counterpart.  
Thus, following Eq.~(\ref{EQ6}), the 
MSD of Si atoms in {\asi} can be expected to be 
somewhat larger than that in {\csi} when an identical 
thermal perturbation is applied to excite 
the system.

\begin{figure}[t!]
\includegraphics[width=0.8\columnwidth]{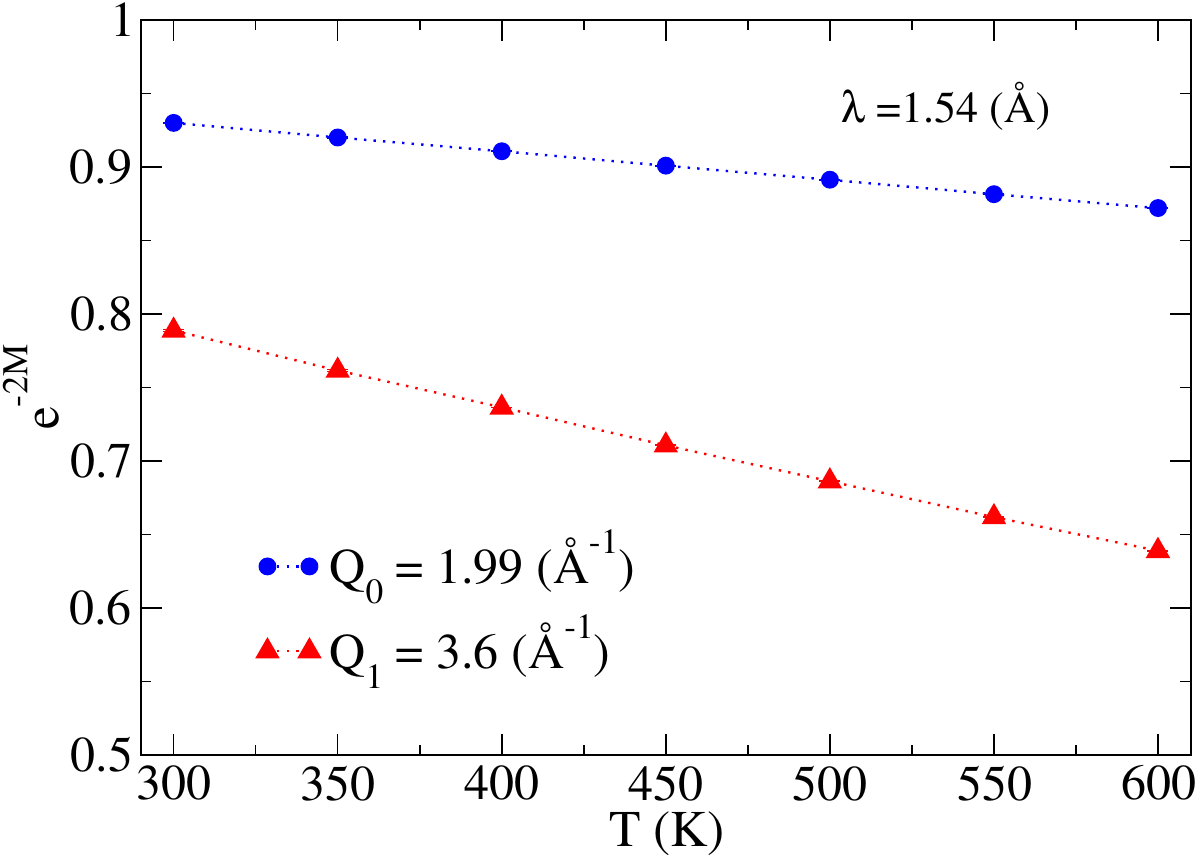}
\caption{
The variation of the Debye-Waller factor of {\asi}, $e^{-2M}$, 
with temperature for $Q$ = 1.99 {\AA}$^{-1}$ (blue circles) 
and 3.6 {\AA}$^{-1}$ (red triangles).  The results are 
obtained by using $\lambda$ = 1.54 {\AA}, which corresponds 
to Cu K$_{\alpha}$ radiation. The $Q$ values chosen correspond 
to the FSDP (1.99 {\AA}$^{-1}$) and the principal peak 
(3.6 {\AA}$^{-1}$) in the structure factor of {\asi}. 
}
\label{F4}
\end{figure}

The temperature dependence of the Debye-Waller factor, 
$e^{-2M}$, is plotted in Fig.~\ref{F4} for $Q$ = 1.99 
{\AA}$^{-1}$ and 3.6 {\AA}$^{-1}$. These two values 
of $Q$ correspond to the position of the first sharp 
diffraction peak (FSDP) and the principal peak in 
the static structure factor of {\asi}, 
respectively. Since the wavevector transfer 
$|Q| = 4\pi \sin(\theta)/\lambda$, the value of the 
Debye-Waller factor at different temperatures for 
the two peaks can be readily calculated for a given 
scattering wavelength $\lambda$. Figure~\ref{F4} shows a plot 
of $e^{-2M}$ versus $T$ for CuK$_\alpha$ X-radiation 
with $\lambda$ = 1.54 {\AA}.  The results suggest that 
the intensity of the principal peak in {\asi} is 
considerably affected by thermal motion of the atoms 
in solids, even at 300 K.  The intensity of the FSDP 
at 300~K, however, is reduced by a factor of about 
0.94, leading to small inelastic scattering at 300 K. The 
missing intensity can be found to be present in 
the background, which originates from temperature-diffuse 
scattering. These results are quite important in comparing 
the static structure factor obtained from the atomic 
coordinates of a computer model to the experimental 
structure-factor data at room temperature, as the 
former does not include any temperature-induced changes 
in the computed data. 

\begin{figure}[t!]
\includegraphics[width=0.8\columnwidth]{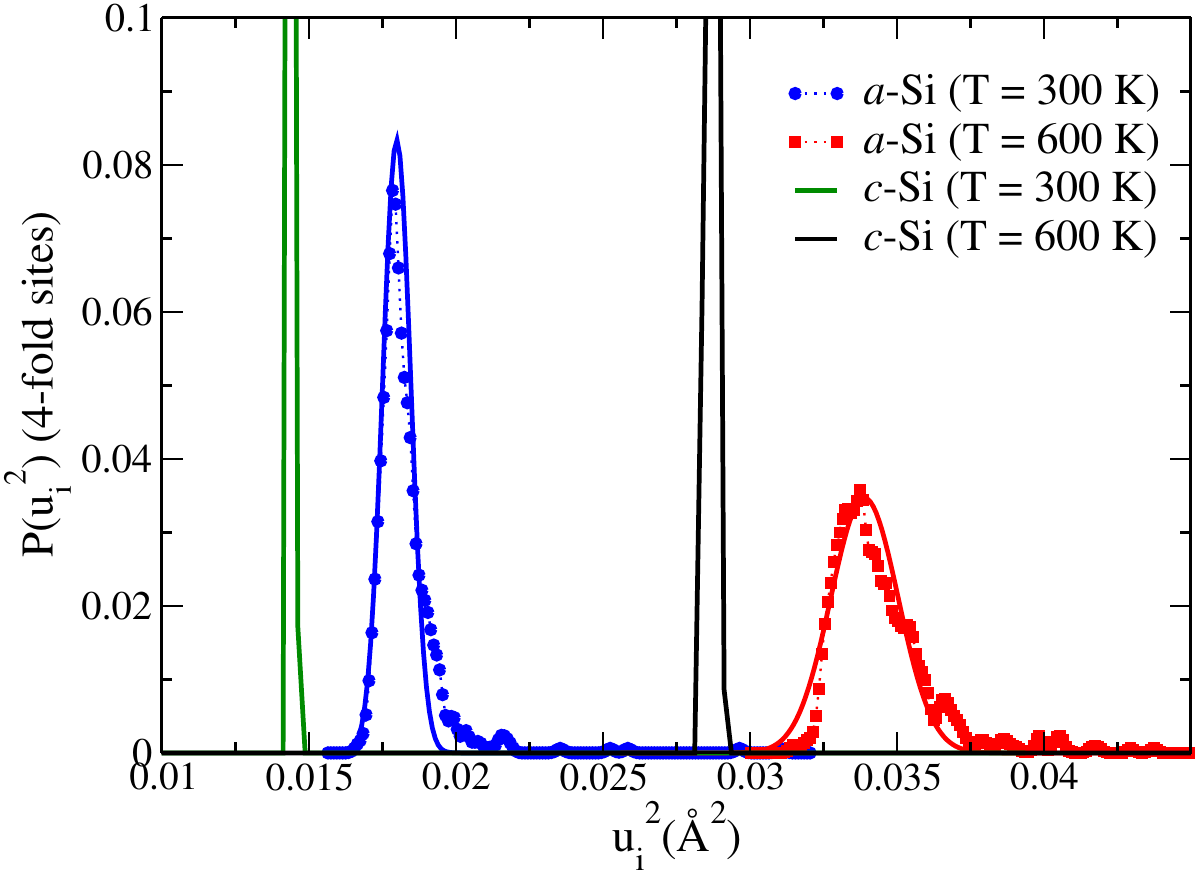}
\caption{
The distributions of local squared displacements, {\ui}, 
for {\asi} at 300~K (blue) and 600~K (red).  The solid 
lines (blue and red) indicate a gaussian fit of the data. 
The corresponding results for diamond-structure {\csi} are also 
shown in the plot (as green and black lines), which are 
truncated for visual clarity and comparison. The 
results for {\asi} correspond to 100\% defect-free 
networks. 
}
\label{F5}
\end{figure}

We now examine the role of short-range ordering on the 
MSD of atoms in amorphous Si networks.  Unlike elemental 
crystals, where one expects a narrow distribution of $u_i^2$, 
induced by thermal vibrations of atoms in an identical 
atomic environment, the MSD of an atom in amorphous 
networks varies from site to site, and it largely 
depends on the local atomic coordination and the 
disorder associated with bond lengths and bond angles. 
Figure \ref{F5} shows the distributions of atomic 
displacements, $P(u_i^2)$ versus $u^2_i$, in {\asi} 
at temperatures 300~K and 600~K, along with their 
crystalline counterparts. The results correspond to 
the average values of {\ui} obtained from two independent 
defect-free configurations of 500 atoms. 
Owing to the crystalline symmetry of atomic positions, 
the distributions for {\csi} in Fig.~\ref{F5} 
are found to be rather narrow and with a root-mean-square 
width that increases with the temperature 
of the system. The heights of these distributions are 
intentionally truncated at a value of 0.1 for 
clarity of presentation and comparison.  
By contrast, the distributions for {\asi} appear 
almost gaussian for small atomic displacements, except for 
a weak non-gaussian tail for high values of {\ui}. 
For crystals, the gaussian shape of $P(u^2_i)$ readily 
follows from lattice-dynamical considerations and it 
can be shown analytically~\cite{Willis:1975} that the 
Fourier transform of $P(u_i^2)$ is directly related 
to the temperature factor of an atom. However, the presence 
of local atomic ordering/disordering in amorphous 
networks can considerably influence the otherwise random 
thermal motion of atoms, leading to a 
notable deviation from an ideal gaussian 
behavior arising from a set of highly disordered sites. 
We shall see soon that this non-gaussian behavior of 
$P(u^2_i)$ is significantly enhanced in the 
presence of coordination defects and other 
inhomogeneities in the network. 

\begin{figure}[b!]
\includegraphics[width=0.8\columnwidth]{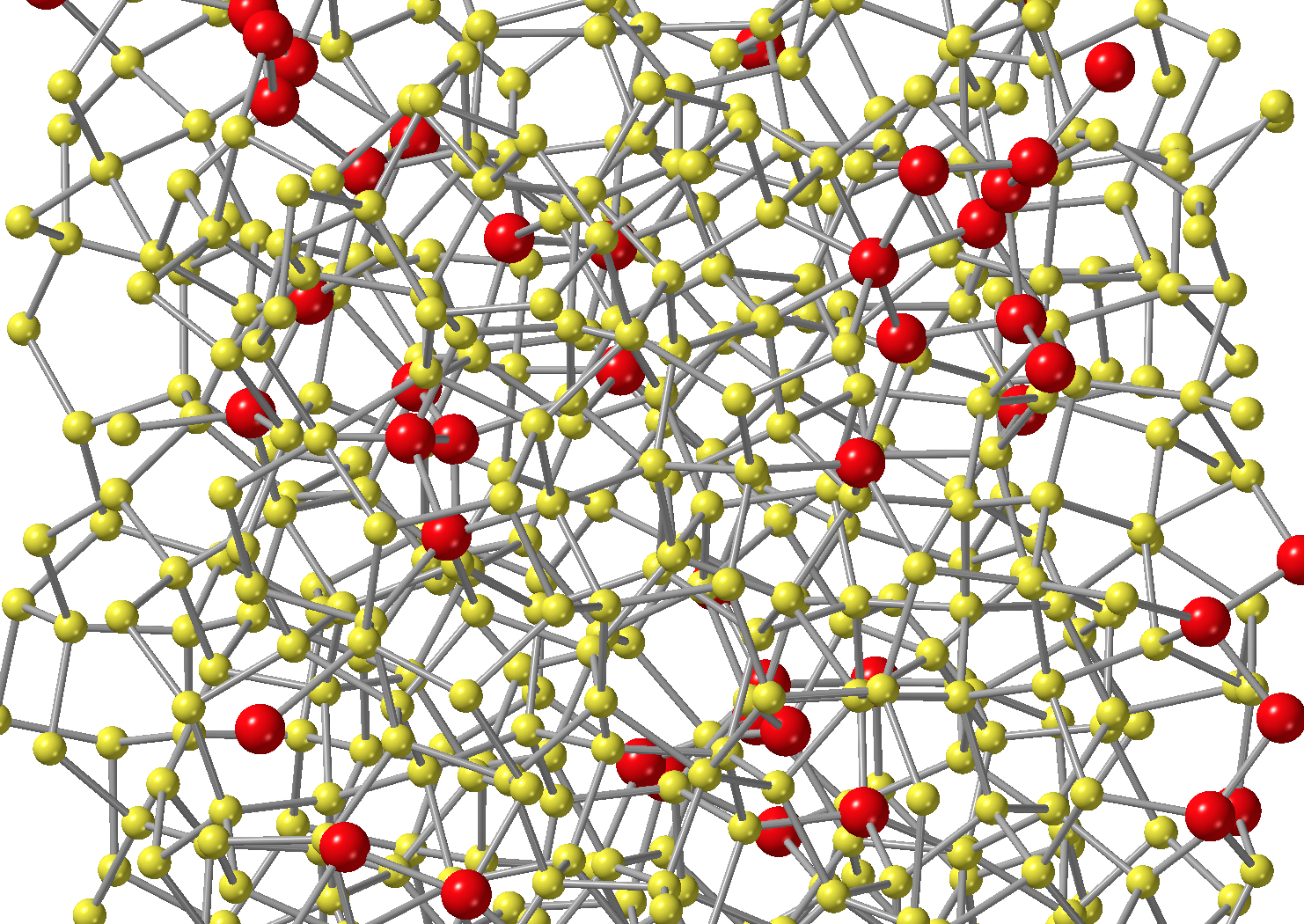}
\caption{
A 500-atom model of {\asi} showing several Si 
atoms with a large value of {\ui} at 300~K.  
Silicon atoms with top 10\% of {\ui} values 
are shown in red color with a slightly larger 
radius.  The rest of the atoms are shown in 
yellow color.
}
\label{F6}
\end{figure}

The origin of the non-gaussian tail, associated 
with high {\ui} values, in defect-free {\asi} 
networks can be traced back to a few clusters 
of Si atoms, which are sporadically distributed 
in the network. These atoms vibrate with 
relatively high amplitudes compared to the 
rest of the atoms in the network. This is 
illustrated in Fig.~\ref{F6}, using a 500-atom model 
of {\asi} at 300 K. Silicon atoms that are 
associated with high {\ui} values, with {\ui} 
$>$ 0.0195 {\AA}$^{2}$, are shown in red color.  
This translates into a value of the atomic displacement, 
which is about 6\% of the average Si--Si bond 
length of 2.36 {\AA}. An examination of the sites 
with high {\ui} values reveals that these sites 
are characterized by the presence of long Si--Si 
bonds, the length of which is about 2--5\% larger 
than the average bond length of 2.36 {\AA}. 
This affects the nearest-neighbor force constants 
($K$) between Si atoms with longer bond lengths 
and results in a reduction of some normal mode 
frequencies (as $\omega^2$ typically decreases 
with decreasing $K$), leading to a larger value 
of {\ui} for these sites from Eq.~(\ref{EQ6}) 
upon thermal excitation. 


\subsection{Effects of defects and inhomogeneities on the MSD}

\begin{figure}[t!]
\includegraphics[width=0.8\columnwidth]{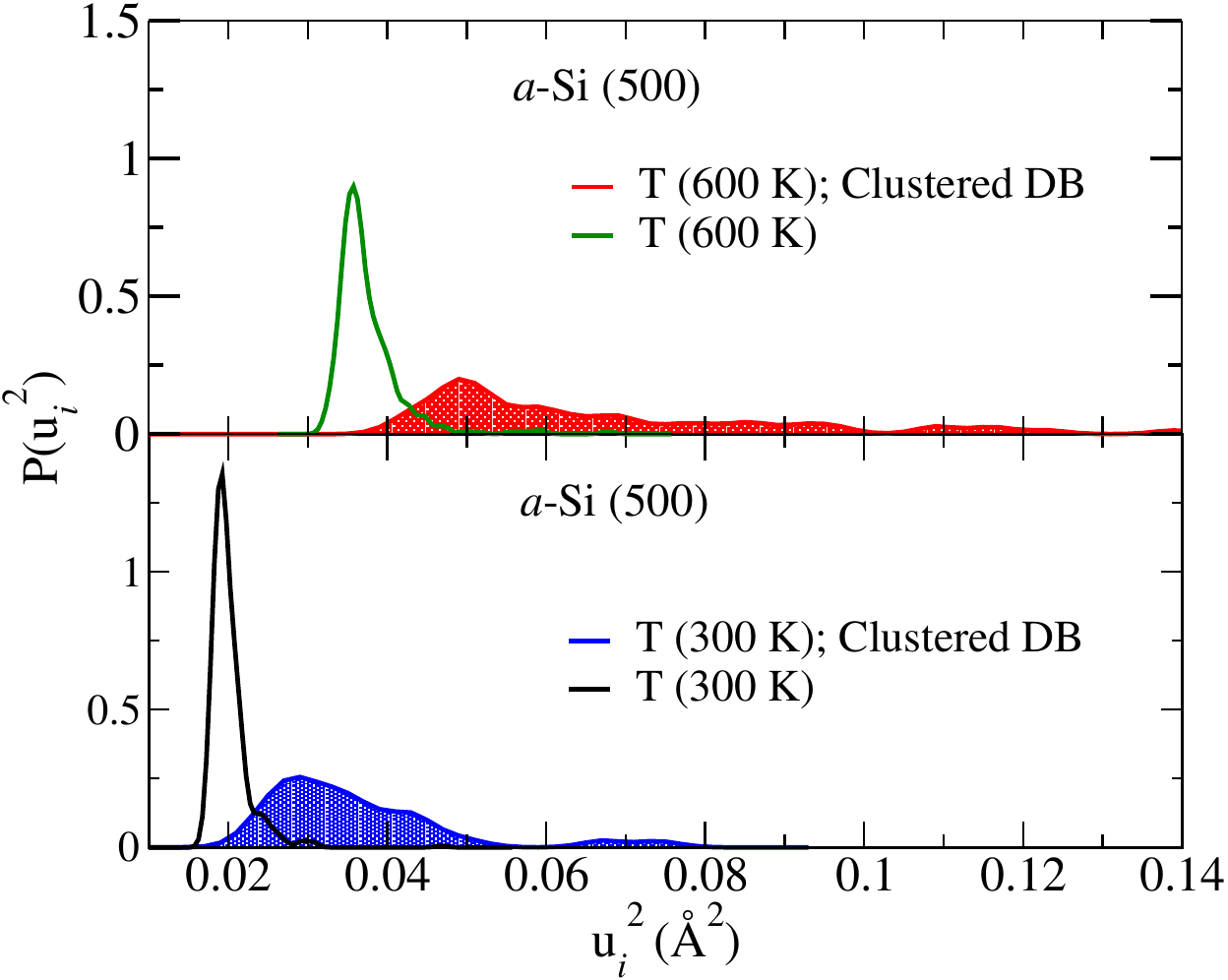}
\caption{
A comparison of the distributions of {\ui} 
for the tetrahedral sites (black and green lines) 
of 100\% defect-free {\asi} networks and 
the clustered DBs (filled blue and red) at 
300~K and 600~K.  The distributions are 
normalized to an integrated value of 
10$^{-2}$.
}
\label{F7}
\end{figure}

\begin{figure}[t]
\includegraphics[width=0.7\columnwidth, angle=270]{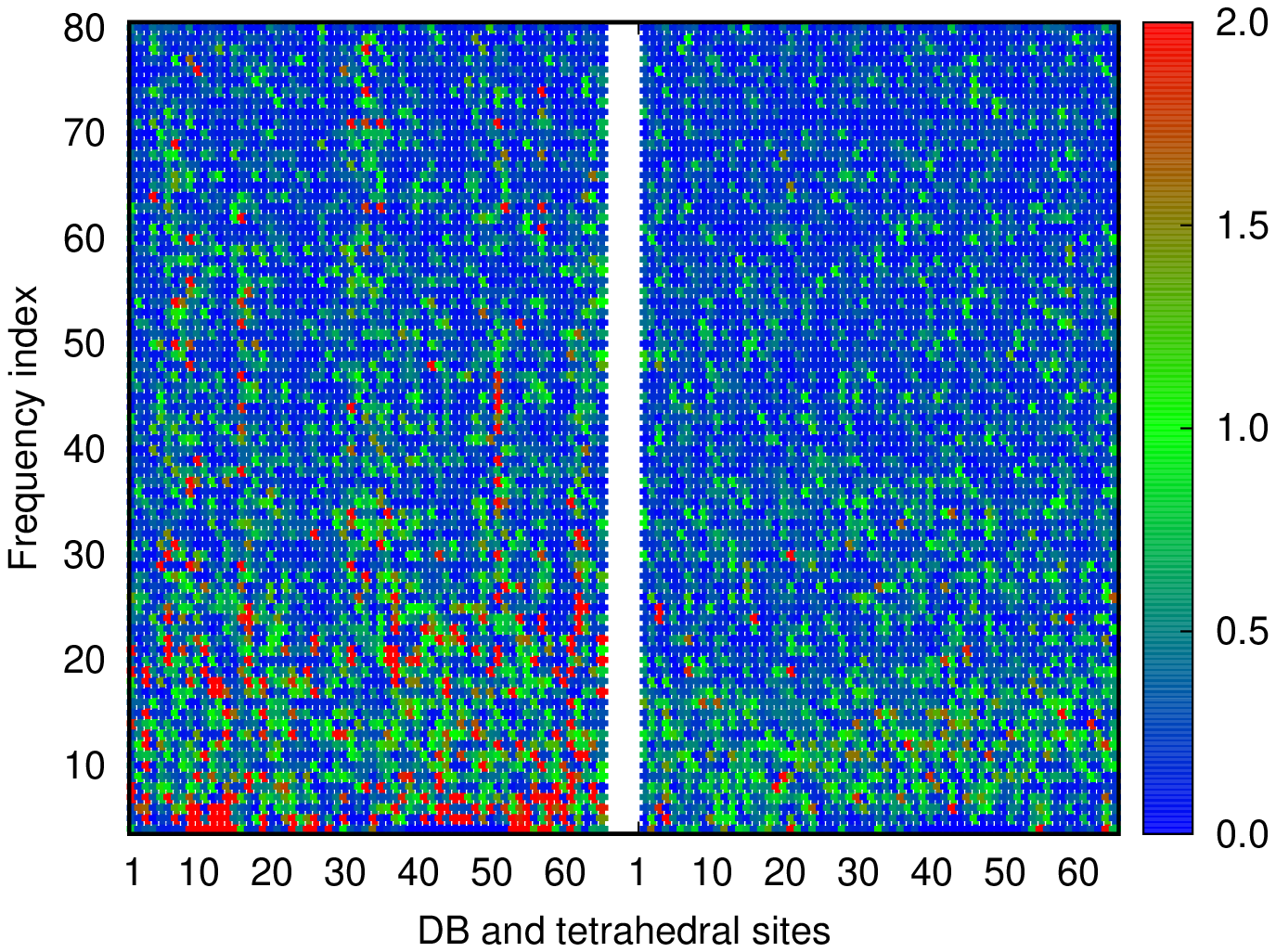}
\caption{ 
A color map showing the contribution to {\ui} at {\em clustered} 
DB sites (left panel) and tetrahedral sites (right panel) from 
the first eighty normal modes (indexed along the Y axis) 
at 300~K,  distributed in the frequency range of 0 and 110 cm$^{-1}$.  
The abundance of red speckles in the left panel indicates a 
large contribution at the DB sites from the low-frequency 
modes. The rightmost vertical bar indicates the percentage 
contribution of $u_i^2 (\nu_j)$ to the total {\ui} from $\nu_j$. 
The results correspond to 65 DB sites and tetrahedral sites 
(indexed along the X axis) selected from the same models. 
}
\label{F8}
\end{figure}

\begin{figure}[t]
\includegraphics[width=0.7\columnwidth, angle=270]{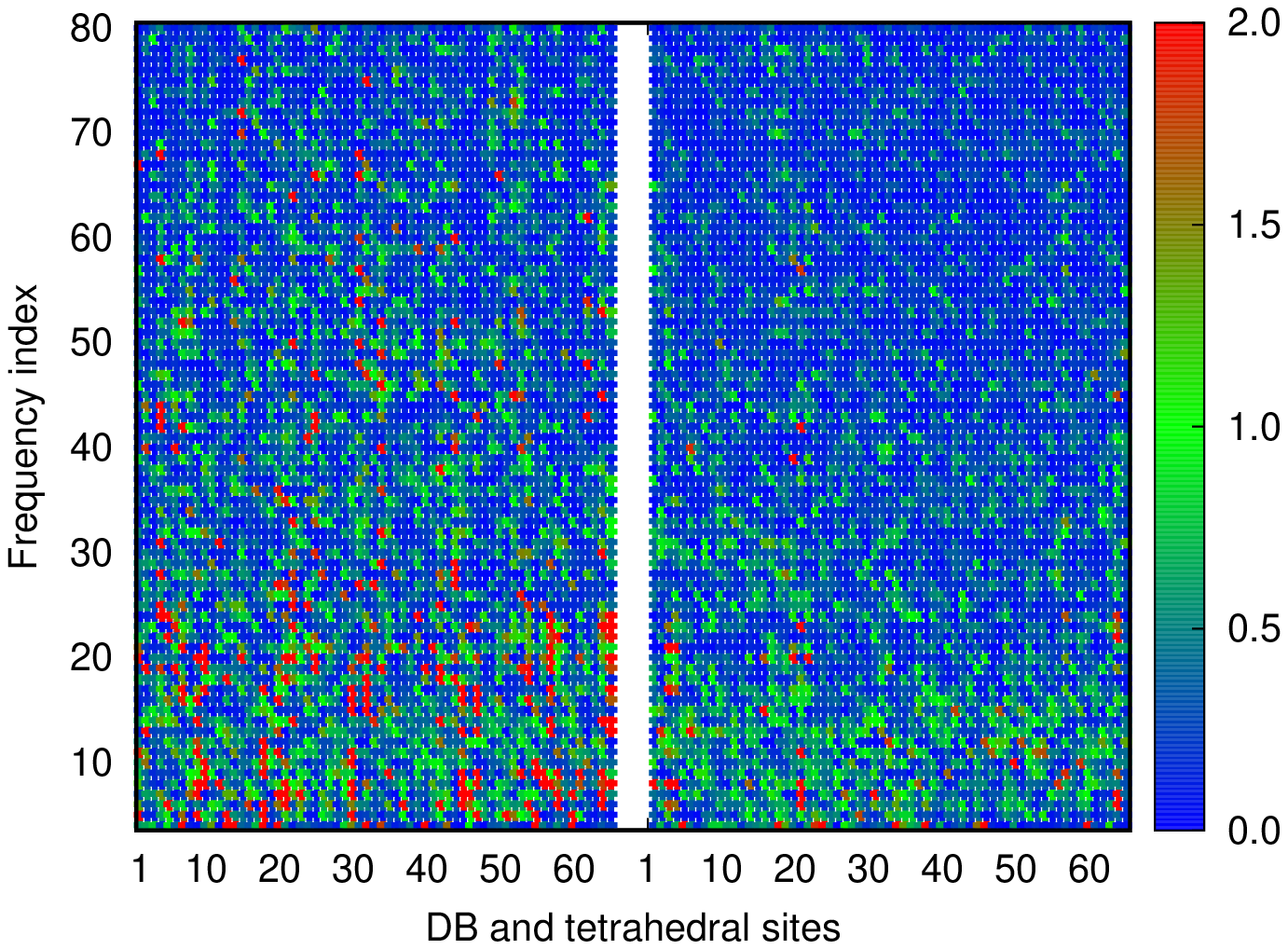}
\caption{ 
A color map showing the contribution to {\ui} at {\em isolated } 
DB sites (left panel) and tetrahedral sites (right panel) from 
individual low-frequency normal modes at 300~K. The left panel 
is awash with red speckles indicating a high contribution from 
the low-frequency in the range of 0--110 cm$^{-1}$ modes at the 
DB sites.  The results correspond to 65 DB sites (left panel) 
and tetrahedral sites (right panel) from the same models. 
}
\label{F9}
\end{figure}

\begin{figure}[t]
\includegraphics[width=0.8\columnwidth]{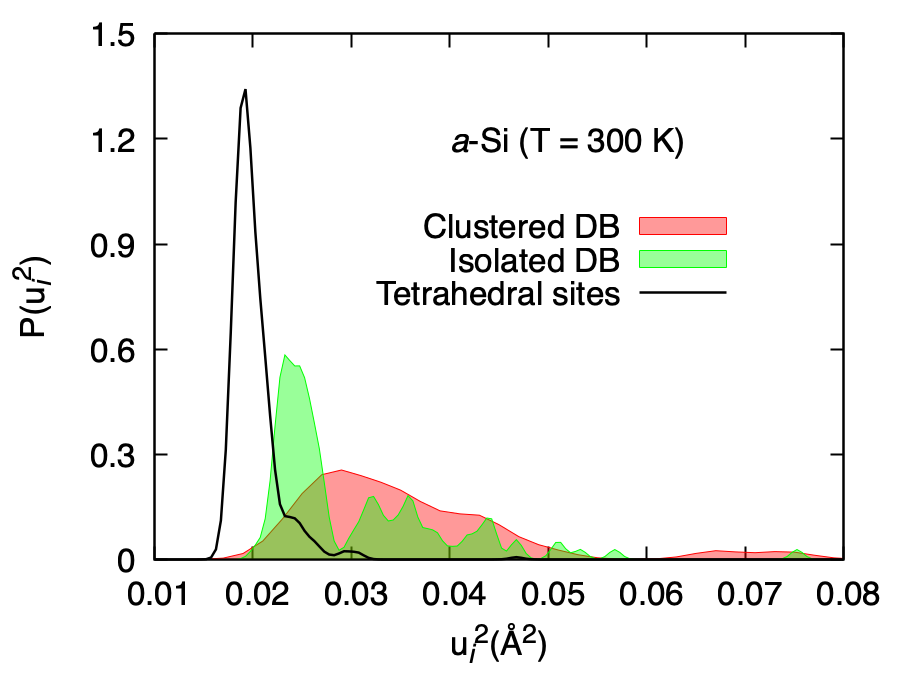}
\caption{ 
The distributions of {\ui} for clustered (red) and  
isolated (green) DBs in {\asi} networks 
at 300~K.  The results correspond to the data 
obtained from 65 DBs from two sets of five 
independent 500-atom models. The results for 
the tetrahedral sites (black)  are from two 
100\% defect-free {\asi} networks. The distributions 
are normalized to an integrated value of 0.01. 
} 
\label{F10}
\end{figure}

Turning now to discuss the role of structural 
defects on the MSD and individual squared 
displacements, we examine the effect of 
three-fold-coordinated atoms, or dangling bonds, 
and extended inhomogeneities, such as voids, 
on the vibrational MSD of atoms.  To this end, 
we have studied two types of dangling bonds. 
The first type of DBs are vacancy induced, 
which were produced by removing a tetrahedrally-bonded 
Si atom, whereas the second type of DBs are 
sparsely distributed in the network. The creation 
of these DBs is discussed in section IIA. 
In the following, we shall refer to these 
dangling bonds as clustered and isolated 
DBs, respectively. 
The results of our calculations, which are 
presented in Figs.~\ref{F7}--\ref{F10}, enable 
us to make the following remarks. 

First, a comparison of {\ui} values obtained for 
the tetrahedral sites with no defects and clustered 
DB sites (in Fig.~\ref{F7}) shows that atomic displacements 
of the DB atoms are notably larger than their 
tetrahedral counterpart by a factor of two or more. 
Second, a small number of DB atoms can be seen to have 
{\ui} values larger than 0.05 {\AA}$^2$ at 300 K, 
and 0.1 {\AA}$^2$ at 600 K.  This indicates substantial 
local movement of some atoms near the defective sites 
and the subsequent healing (or reconstruction) of a 
dangling bond to form a tetrahedrally-bonded Si atom. 
Indeed, it was observed that a considerable number of clustered DBs 
introduced in the network reorganized themselves to 
form stable tetrahedral bonding following thermalization 
and structural relaxation. 
Third, a relatively large value of {\ui} at the DB 
sites can be partly attributed to the presence of 
reduced atomic coordination, which renders the atoms 
more susceptible to move. 
A normal-mode analysis reveals that a few low-energy 
modes contribute considerably to {\ui} of the DBs 
compared to the tetrahedral sites. This is evident 
from Fig.~\ref{F8}, where a color-map representation 
of $u_i^2(\nu_j)$ -- the contribution to {\ui} 
at the DB site $i$ from the normal mode $\nu_j$ --  is 
presented against the normal-mode frequencies (along the Y axis) 
and atomic sites (along the X axis). 
The left panel in Fig.~\ref{F8} corresponds to the 
results obtained for the clustered DBs at 300~K, 
whereas the right panel shows the same for an 
identical number of tetrahedrally-bonded atoms. 
Atomic indices of sixty-five clustered/tetrahedral 
sites are indicated along the abscissa, whereas 
the first eighty low-frequency normal modes are 
indicated along the ordinate.  The color in the 
plot is indicative of the partial contribution of 
$u_i^2(\nu_j)$ (in percent) to the total MSD 
of atom $i$. The significant presence of red 
speckles in the left panel is indicative of high 
contributions arising from a few tens of low-frequency 
normal modes at the clustered DB sites. The vertical 
indices along the Y axis correspond to the frequency 
range of 0 to 110 cm$^{-1}$ for $j$=1 to 80. 
A similar observation applies to Fig.~\ref{F9} 
where the results for the isolated DBs are 
presented. 

The distributions of {\ui} obtained for the 
clustered and isolated DBs are found to be quite 
different from each other.  This is evident from 
Fig.~\ref{F10}, where the distributions resulting
from the clustered (filled red) and isolated (filled green) 
DBs at 300~K are plotted.  The fine structure in the distribution 
for the latter is reflective of the degree of 
sparsity and the disorder associated with the 
isolated DBs in the network.  
The first green peak in Fig.~\ref{F10} indicates 
the presence of several truly isolated DBs 
in the network. The subsequent green peaks 
are reflective of a somewhat lesser degree 
of sparsity of the remaining isolated DBs in the 
network. 
An analysis of the distribution of the isolated 
DBs in the networks shows that almost half of 
the 65 DBs are sparsely distributed in the network 
with an average separation distance of 10.3 {\AA}, 
and a good majority of these sites contribute to 
the first (green) peak.

\begin{figure}[t]
\includegraphics[width=0.8\columnwidth]{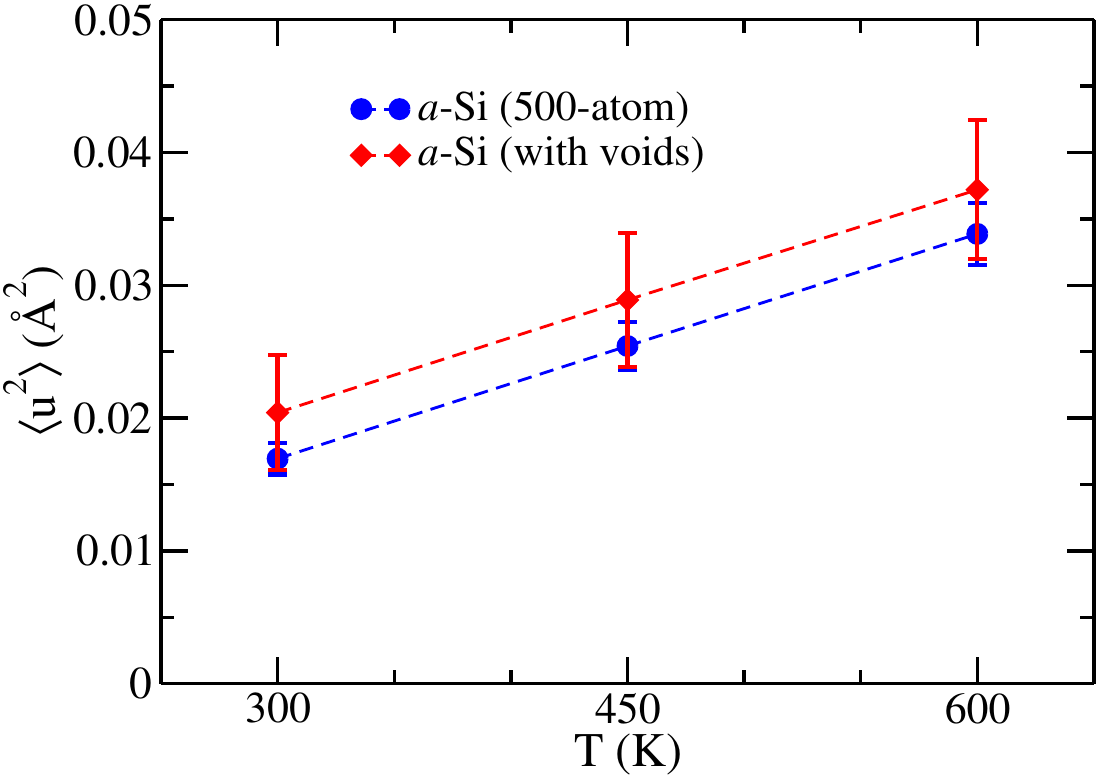}
\caption{
The MSDs of Si atoms obtained from 500-atom models 
of {\asi} in the presence of two voids of diameter 
8~{\AA} (red diamonds) at 300~K, 450~K, and 600~K. 
The corresponding values of the MSD without voids 
(blue circles) are also shown in the plot for 
comparison. 
} 
\label{F11}
\end{figure}

Likewise, the effect of microvoids on the MSD of 
atoms can be studied by introducing a couple of 
voids in the network. Figure \ref{F11} shows 
the MSD at 300 K, 450 K and 600 K, before 
and after introducing two voids of diameter 8 {\AA} 
in two 500-atom models of {\asi}. The plots in 
Fig.~\ref{F11} show that the MSD has considerably 
increased due to the presence of several
defective atoms on the surface of the voids. The 
presence of local disorder and reduced coordination 
considerably weakens the effective force constants 
between neighboring atoms that lead to an increase 
of the MSD of the atoms on the surface of the voids. 

\subsection{Debye temperature of {\asi} from the MSD of atoms}

Having discussed the variation of the MSD of atoms with 
temperature in the presence of disorder and defects, we 
now obtain estimates of the Debye temperature, $\Theta_d$, 
and compare the results with those from experiments. 
To this end, we first note the following assumption about 
the definition of the Debye temperature for amorphous 
solids. In experiments, 
the Debye temperature of crystalline solids is generally 
determined by comparing the measured 
specific-heat data at low temperatures (typically below 30 K) with 
the expression for the specific heat, $C_v$ = $aT + bT^3$, 
in the Debye approximation, where the coefficient $b$ involves 
the Debye temperature $\Theta_d$. 
At sufficiently low temperatures, the contribution to 
$C_v$ originates from the (distribution of) 
low-frequency vibrational modes in crystals, which 
can be adequately represented by a quadratic function 
of the frequency. This results in a $T^3$ dependence 
of the specific heat in crystals. As a result, one can 
justify the use of the Debye approximation and the 
resulting Debye temperature as a physical parameter 
for crystalline solids. For 
amorphous solids, however, the above reasoning is 
weakened by the presence of excess low-frequency 
vibrations (near 1-2 THz) in many systems, leading 
to a vibrational density of states that may not be 
well represented by a quadratic function of frequency 
in the low-frequency region.  This can affect the 
temperature dependence of the specific heat of 
amorphous solids. 
It has been shown~\cite{Pohl:1981} that for a number 
of glassy systems, the coefficient $b$ can be greater 
than the Debye coefficient $b_d$ even if the $T^3$ 
dependence of the specific heat is assumed to hold 
at low temperatures. 
Since the experimental values of $\Theta_d$ of {\asi} reported 
in the literature~\cite{Zink:2006,Mertig:1984} are 
obtained by assuming $b = b_d$, we shall make the same 
assumption to calculate $\Theta_d$ for the purpose of 
comparison. 

To calculate a theoretical estimate of the Debye temperature 
of {\asi}, we adopt here two distinct approaches. The first approach 
involves the use of the computed values of the MSDs at low 
temperatures by comparing with those obtained from the Debye 
theory. This can be achieved by writing the expression for the 
MSD at temperature $T$ in the Debye 
approximation~\cite{Willis:1975}

\be 
u^2_d(T) = \frac{9\hbar^2}{mk_BT} 
\left[\frac{1}{4x_d} + \frac{1}{x^3_d} \int_0^{x_d} \frac{x\, dx}{e^x - 1} \right], 
\label{EQ11} 
\ee 
where $x_d = h\nu_d/k_BT = \Theta_d/T$ and $\nu_d$ is 
the Debye frequency.  In our approach, we proceed to 
calculate $\Theta_d$ from Eq.~(\ref{EQ11}) by replacing 
$u^2_d(T)$ with $\langle u^2(T)\rangle$. 
To ensure that the Debye approximation remains valid, we 
limit ourselves to $\langle u^2(T)\rangle$ values 
obtained at low temperatures of up to 50 K from 
two 500-atom models.
The Debye temperature is obtained numerically by computing 
the value of $x_d$ for which Eq.~(\ref{EQ11}), with $u^2_d(T) 
\to \langle u^2(T)\rangle$, is satisfied at temperatures of 
10 K, 30 K and 50 K, and the average value of $x_d$ and $\Theta_d$ 
are computed from the results. Table \ref{tab1} lists 
the results from our calculations.  
The estimated value 
of the average Debye temperature, $\langle \Theta_d \rangle $, 
and the corresponding value of the average Debye frequency, 
$\nu_d$, for {\asi} are found to be 541.5$\pm$4~K and 11.3 THz, 
respectively, over a temperature range of 10--50~K. 
We should mention that the absence of low-frequency 
vibrational modes in small 500-atom {\asi} models may 
underestimate the MSD values at low temperatures due to the 
inverse-square dependence of the MSD on the frequency (cf.~Eq.~\ref{EQ6}). 
Since the $x_d$  value in Eq.~(\ref{EQ11}) is found to 
decrease with an increasing value of $\langle u^2(T)\rangle$~\cite{U2},  
the value of $\Theta_d$ obtained here from 500-atom models 
provides an upper bound of the Debye temperature of {\asi}.

The computed values of $\Theta_d$ obtained from the MSDs at 
three different temperatures are presented in table \ref{tab1}. 
These values are  found to 
be considerably higher than the theoretical value of 
430--470~K obtained by Feldman et al~\cite{Feldman1991}. 
The latter employed 216-atom models of {\asi} and 
classical potentials to compute the elastic constants of {\asi} 
and hence the Debye temperature. The difference between 
these two sets of values is not unexpected as the computation 
of $\Theta_d$ from different theoretical approaches 
may vary notably, depending upon model sizes, the 
structural quality of models, and the accuracy of 
total-energy and forces used in the calculations. 
On the other hand, experimental values of $\Theta_d$, 
obtained from specific-heat measurements of {\asi} at 
low temperatures in the Debye approximation, suggest 
that $\Theta_d$ varies from 487$\pm$5~K~\cite{Zink:2006} 
to 528$\pm$20 K~\cite{Mertig:1984}.  The value 
obtained from the inversion of Eq.~(\ref{EQ11}) is quite 
close to that of Mertig et al.~\cite{Mertig:1984} but 
notably higher than that observed by Zink et 
al.~\cite{Zink:2006} recently. In the next section, we shall 
further address this issue by calculating the specific 
heat of {\asi} at low temperatures and obtaining a 
value of $\Theta_d$ from the Debye approximation for 
amorphous solids. 

\begin{table}[t] 
\centering
\caption{
Debye temperature ($\Theta_d$) of {\asi} estimated from 
the mean-square displacement (MSD) of atoms at the 
temperature range of 10--50~K . $\langle\Theta_d\rangle$ 
indicates the average Debye temperature for this range.
}
\vskip0.2cm
\begin{ruledtabular}
\begin{tabular}{c|c|c|c|c|c|c}
T & $\langle u^{2}$ $\rangle$ & $x_{D}$ & $\Theta_{d}(T)$ & $\langle \Theta_d \rangle$ &$\Theta_{d}$ (Expt.) & $\Theta_{d}$ (Elastic) \\%& $\theta_{D}^{\footnote{From Refs.~\cite{Eguchi:1987,Batterman:1962,Reichelt:1966}}} $ \\
(K) & ({\AA}$^{2}$) & & (K) & (K) & (K) \\
\hline
10\: &  0.00725 & 53.632 & 536.31  &        &  & \\
30\: & 0.00730   & 18.092 & 542.76 & 541.5  & 487$\pm$ 5$^{\footnote{From Refs.~\cite{Zink:2006}}}$ & 430-470$^{\footnote{From Refs.~\cite{Feldman1991}}}$\\
50\: &   0.00752 & 10.903 & 545.17 &  & 528$\pm$ 20$^{\footnote{From Refs.~\cite{Mertig:1984}}}$ & \\
\end{tabular}
\end{ruledtabular}
\label{tab1}
\end{table}

\subsection{Debye temperature of {\asi} from specific-heat calculations}

The quantum-mechanical calculation of the molar specific heat ($C_v$) 
of {\asi} is rather straightforward and has been 
reported by a number of workers in recent 
years~\cite{Limbu:2019,Igram:2018,Queen:2013}.
In {\it ab initio} lattice-dynamical calculations 
in the harmonic approximation, $C_v$ can be obtained 
from 

\be 
\frac{C_v}{3R} = \frac{1}{3N}\sum_{i=1}^{3N} 
\frac{k_b\,x_i^2\, e^{x_i}}{(e^{x_i} - 1)^2}, \: \: 
\mbox{where} \:  x_i = \frac{\hbar\omega_i}{k_bT}, 
\label{EQ12} 
\ee 
and the corresponding Debye expression is given by 
\be
\left[\frac{C_v}{3R}\right]_d = \frac{3}{x_d^3} \int_0^{x_d} 
\frac{x^4 e^x\, dx}{(e^x - 1)^2} \; \; \text{with} \; x_d = \Theta_d/T. 
\label{EQ13}
\ee

However, the calculation of the Debye temperature from a 
$C_v$-vs-$T$ plot for small models of {\asi} 
is highly nontrivial as it requires the calculation of $C_v$ 
values at very low temperatures for which the major 
contribution to $C_v$ comes from the low-frequency 
vibrational modes. Since these low-frequency 
modes (with $\hbar\omega \approx k_bT$ for $T$ = 10--30 K) 
cannot be realized/formed in small finite-size models and 
the full self-consistent field density-functional 
calculation of the vibrational modes for large models 
using extended basis states is computationally intractable, 
it is extremely difficult 
to obtain a realistic estimate of $\Theta_d$ from small 
models in {\it ab initio} calculations. 
Below, we illustrate this point by studying results 
from 500-atom and 2000-atom {\asi} models.  

\begin{figure}[t!]
\includegraphics[width=0.4\textwidth]{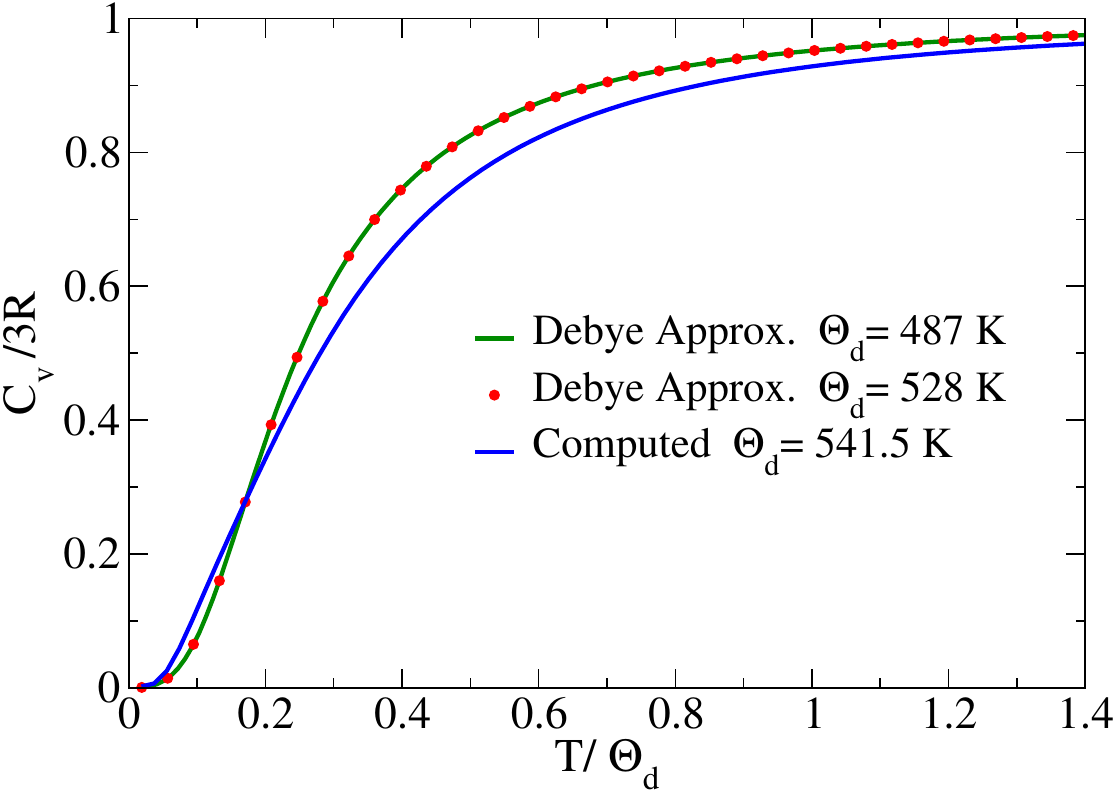}
\caption{
The temperature dependence of the molar specific heat, $C_v$, 
of {\asi} from the Debye approximation (green line and red 
circles), and {\it ab initio} lattice-dynamical calculations 
(blue line) using 500-atom {\asi} models. The results for 
the Debye approximation are obtained by using the experimental 
values of $\Theta_d$ = 528~K and 487~K, whereas the average 
value of $\Theta_d$, 541.5~K (from table \ref{tab1}), is 
used for scaling the X~axis to present the results from the 
lattice-dynamical calculations.  
}
\label{F12}
\end{figure}

\begin{figure}[ht!]
\includegraphics[width=0.4\textwidth]{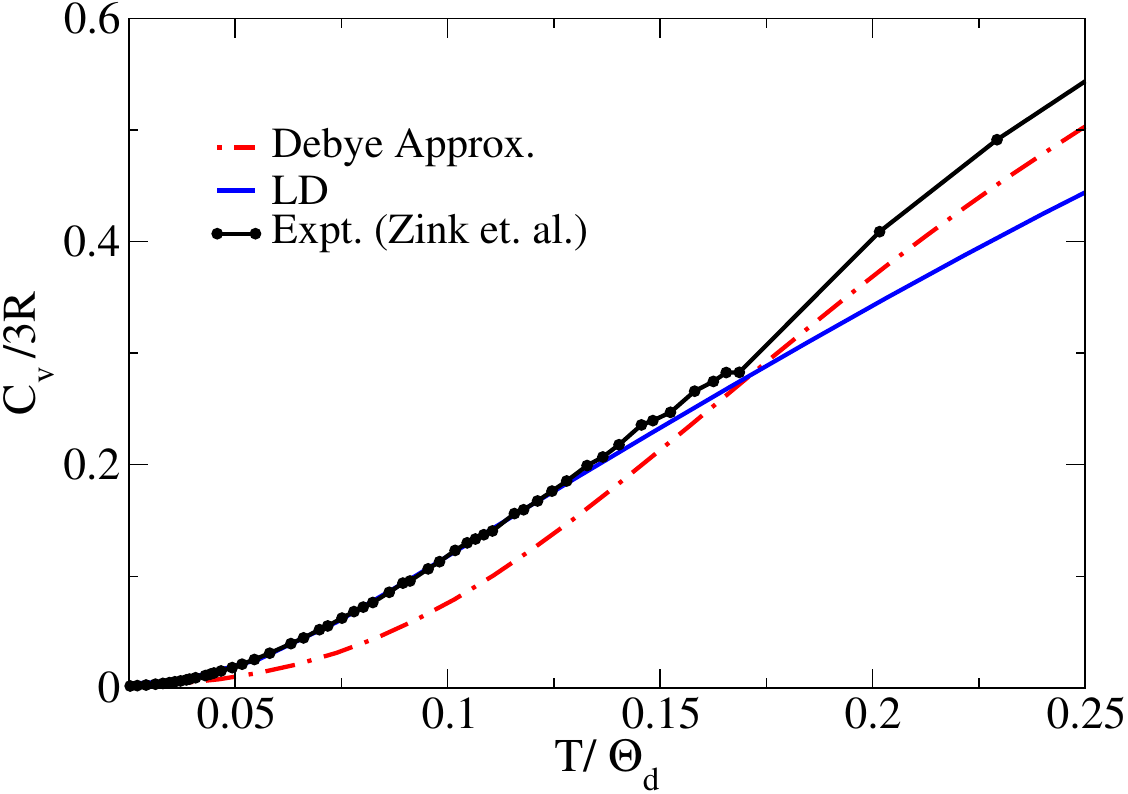}
\caption{
Low-temperature behavior of $C_v$ in {\asi} from the Debye 
approximation (red line), {\it ab initio} lattice-dynamical (LD)
calculations (blue line), and experiments~\cite{Mertig:1984} 
(black line) in the region of 20--135~K. An approximate 
linear behavior of $C_v$ can be seen to appear in the 
vicinity of 60 K in both experimental and theoretical results, 
which corresponds to a frequency value of about 1.25 THz (for 
$\Theta_d$ = 541.5~K).
}
\label{F13}
\end{figure}

\begin{figure}[t!]
\includegraphics[width=0.4\textwidth]{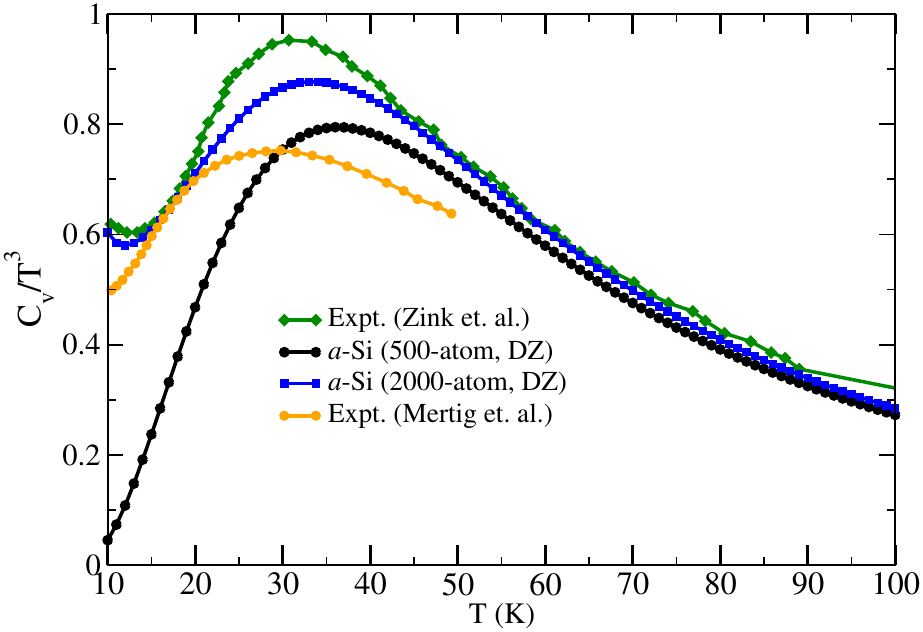}
\caption{
Comparisons of the specific heat, $C_v$, of {\asi} at 
low temperatures from experimental data (green~\cite{Zink:2006} 
and orange~\cite{Mertig:1984}) and the present study, 
using 500-atom (black) and 2000-atom (blue) models. 
The computed values of $C_v$ are obtained from using the 
double-zeta (DZ) basis functions and the GGA in 
{\it ab initio} lattice-dynamical calculations. 
}
\label{F14}
\end{figure}

The results from Eqs.~(\ref{EQ12}) and (\ref{EQ13}) 
are plotted in Figs.~\ref{F12}--\ref{F13}, and Fig.~\ref{F14} 
as a function of $1/x_d$ and $T$, respectively. An 
analysis of the plots leads to the following observations: 

1) A small change of $\Theta_d$, and hence $x_d$, has 
very little to no effect on $C_v$ vs. $T/\Theta_d$ plots. 
This is not unexpected in view of the integral 
nature of Eq.~(\ref{EQ13}) and it is reflected in 
Fig.~\ref{F12}, where $\Theta_d$ = 487~K and 528~K 
were used to obtain $C_v$ in the Debye approximation, 
leading to almost identical values of $C_v$. 
It is evident from Fig.~\ref{F12} that the Debye 
approximation considerably overestimates the 
value of $C_v$ in {\asi} -- compared to that from 
{\it ab initio} lattice-dynamical calculations -- above 
a certain temperature $T_c$ ($\sim$ 92 K) and slightly 
underestimates it below $T_c$. Figure~\ref{F13} 
zooms in on the low-temperature region of $C_v$, 
which clearly indicates that the crossover temperature 
$T_c$ is in the vicinity of 0.17$\Theta_d$ or about 
92~K for $\Theta_d$ = 541.5~K; 

2) The linear variation of $C_v$ with $T$ in the 
low-temperature region of 40--80~K (i.e., $x_d^{-1} 
\approx$ 0.07 to 0.15) in Fig.~\ref{F13} markedly 
deviates from the Debye-$T^3$ law at low temperatures. 
A linear behavior of $C_v$ is well-known in the experimental literature of 
non-crystalline solids, including for {\it a}-SiO$_2$, 
{\it a}-Se~\cite{Zink:2006,Zeller:1971,Hornung:1969} 
and dilute magnetic alloys~\cite{Marshall:1960}, 
at temperatures of up to 10 K. However, this behavior is often 
attributed to the presence of tunneling modes in 
two-level systems (TLS)~\cite{Anderson1972} at very 
low temperatures or due to the presence of vibrational modes in reduced 
dimensions, where the presence of ordered/disordered 
parallel atomic chains (e.g., in Se/{\it a}-Se) can lead to 
a linear behavior over a certain range of temperature. 
By contrast, the linear variation of $C_v$ observed here 
in Fig.~\ref{F13} for {\asi}, which is evident in both 
experimental and theoretical results, appears at 
considerably higher temperatures of 40-80 K than 10 K. 
This observation is quite remarkable as we are not 
aware of any earlier theoretical works on {\asi} 
that demonstrate this experimentally observed linear 
behavior of $C_v$ with $T/\Theta_d$ at temperatures 
of up to 80~K; 

3) The peak positions in the experimental and computed 
data for $C_v/T^3$ vs.\,T plots in Fig.~\ref{F14} are found 
to be at 28--30~K and 33--35~K, respectively. A small rightward 
shift of the computed peak for the 500-atom model with 
respect to its experimental counterpart plausibly arises 
from the absence of some low-frequency vibrational modes in the 
model. This also explains a considerable reduction 
of $C_v$ values at low temperatures (below 50~K); the system 
needs a relatively less amount of energy to excite the few 
low-frequency modes that are present in small 500-atom models.  

In view of the preceding observation, it is appropriate to conclude 
that the $C_v$ values obtained from small computer-generated 
models cannot accurately reproduce the experimental $C_v$ data 
of {\asi} below 50~K, and hence the Debye temperature which is obtained 
from the low-temperature $C_v$ data in the Debye 
approximation. This is evident in Fig.~\ref{F14}, where the results from 
the lattice-dynamical calculations for 500-atom and 
2000-atom {\asi} models are compared with experimental data. The difference 
between the experimental and theoretical values of $C_v$ 
below 50~K is considerable for 500-atom models but the 
results improved significantly as the model size increases 
to 2000 atoms. 
Finkemeier and von Niessen~\cite{Finkemeier:2000} 
have shown using classical calculations that the size of 
{\asi} models should be of the order of a few tens of thousands 
of atoms in order to produce the so-called excess low-frequency 
modes near 1~THz~\cite{fv}. 
This possibly explains the observed deviation of theoretical 
$C_v$ values from experimental data in the region of 
20--40~K in Fig.~\ref{F14} for 2000-atom models. A 
comparison of the experimental data by Zink et al.~\cite{Zink:2006} 
with those from 2000-atom models in the temperature range 
of 10~K to 20~K suggests that the theoretical value of 
the coefficient $b$ ($\approx 0.58$) is quite close to the 
experimental value of 0.6, leading to a theoretical value 
of the Debye temperature of 488 K.

\subsection{
Effects of anharmonicity on the MSD at high temperatures
}
We now briefly discuss the effects of anharmonicity 
on the MSD of atoms in {\asi} at high temperatures. 
It goes without saying that the temperature-induced 
anharmonic effects at high temperatures can be 
truly taken into account by including thermal expansion 
of solids in simulations. Since the AIMD simulations presented 
in section~IIC were conducted in canonical and microcanonical 
ensembles, where the volume expansion of {\asi} was 
taken into account on an ad hoc basis, 
it is not possible to accurately address 
the role of anharmonicity on the MSD of atoms. Nonetheless, 
the results from direct AIMD simulations should provide 
a glimpse of the anharmonic effects on the MSD 
at high temperatures, as the volume-expansion factor, 
$\sim (1 + 3\gamma\Delta T)$, in Eq.~(\ref{EQ9}) turns out 
to be very small and of the order of 1.0036 for 
$\Delta T$ = 300~K and $\gamma = 4 \times 
10^{-6}$~K$^{-1}$~\cite{Takimoto}. Thus, if we make allowances for not 
including such a small change of volume of the system, 
we should be able to observe the effects of temperature-induced 
anharmonicity in the potential, if present, on the MSD of 
atoms at high temperatures obtained from the direct 
AIMD results. 

\begin{figure}[t!]
\includegraphics[width=0.8\columnwidth]{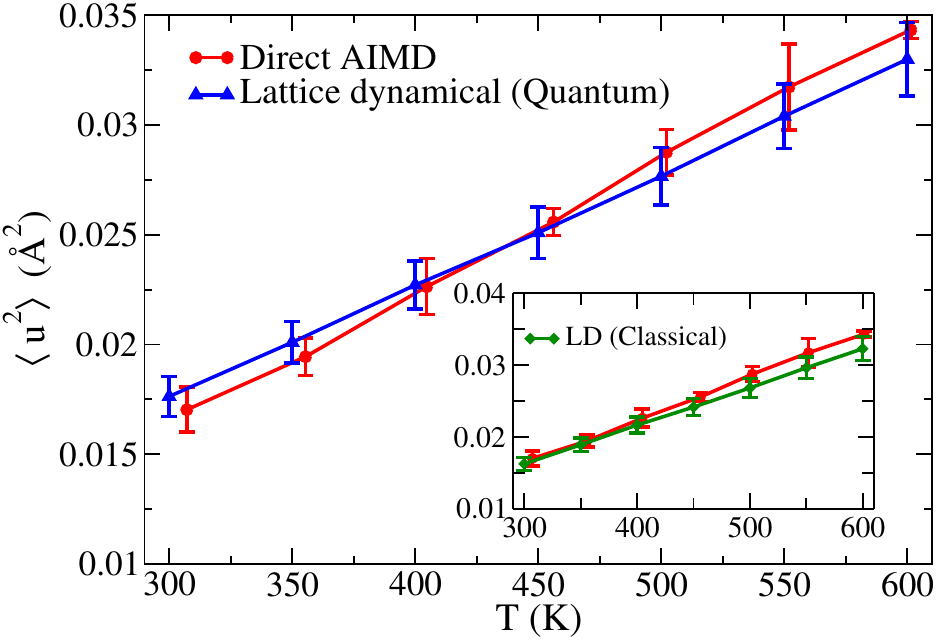}
\caption{ 
A comparison of {\u2} obtained from 216-atom models
of {\asi} using lattice-dynamical calculations 
(blue) and direct AIMD 
simulations (red). The inset shows the 
lattice-dynamical results obtained by using 
the classical expression of energy, $\langle E_n \rangle = k_BT$ 
in Eq.~(\ref{EQ6}).  
}
\label{F15}
\end{figure}

Figure \ref{F15} shows the results obtained from 
216-atom models of {\asi} using direct AIMD simulations 
in NVT and NVE ensembles, along with the results from 
lattice-dynamical calculations for an identical 
216-atom model in the harmonic approximation. The results enable us to make the 
following remarks. First, a comparison of {\u2} obtained 
from the lattice-dynamical calculations with those 
from direct AIMD simulations suggests that anharmonicity 
does not play a significant role at temperatures below 450~K. 
A small difference between two sets of data below 450~K 
can be attributed to the use of the phonon distribution 
function, $1/[\exp(x)-1]$ with $x = \hbar\omega/k_BT$, in 
the lattice-dynamical calculations. This is apparent from 
the plot shown as an inset in Fig.~\ref{F15}, where the 
lattice-dynamical results are re-calculated in the 
classical limit using the value of $\langle E_n \rangle = k_BT$ 
in Eq.~(\ref{EQ6}). Since the AIMD simulations are conducted 
by integrating the classical equations of motion, the 
results obtained from equilibrium atomic trajectories 
using Eq.~(\ref{EQ10}) are reflective of the classical equipartition 
theorem. Consequently, the direct AIMD results are somewhat 
underestimated in the temperature range of 300--400~K. 
Second, the deviation at high temperatures, above 450~K, 
is likely to originate from anharmonic effects in the 
dynamics. This is partly due to large thermal vibrations, 
which can affect the linear dependence of the forces 
on atomic displacements, and in part due to 
disorder in the amorphous network affecting force 
constants between neighboring atoms.  
Since the experimental value of the Debye temperature 
in {\asi} is about 487--528~K, one may assume that the MSD 
of atoms can be calculated fairly accurately in the 
classical limit from the direct AIMD trajectories at 
temperatures above 450~K. Lastly, the use of the 
volume-expansion factor of $(1+3\gamma\Delta T)$ in 
our direct AIMD simulations does not yield any noticeable changes of 
the MSD of atoms, even at 600~K. This observation 
is consistent with the results observed in 
Fig.~\ref{F1}.  It is therefore reasonable to conclude 
that the observed deviation of the direct AIMD 
results from its lattice-dynamical counterpart at 
temperatures above 450~K can only result from the 
presence of a weak anharmonic part in the Si-Si potential 
at high temperatures of up to 600~K. 

\begin{figure}[t!]
\includegraphics[width=0.4\textwidth]{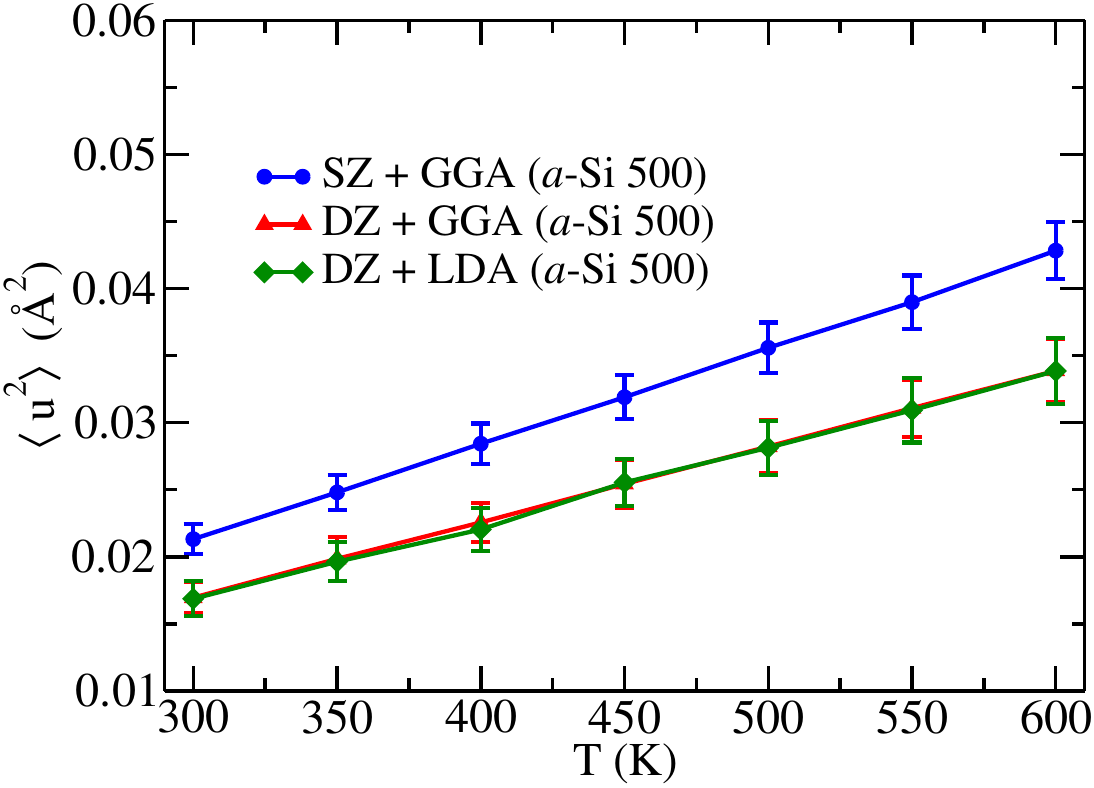}
\caption{
The dependence of the MSD, {\u2}, on the single-zeta (SZ)
and double-zeta (DZ) basis functions and the 
exchange-correlation (XC) approximations (LDA and GGA) 
used in the DFT calculations in this study.  
The data correspond to the average values obtained 
at each temperature from two independent 500-atom 
configurations. 
}
\label{F16}
\end{figure}

\begin{figure}[t!]
\includegraphics[width=0.4\textwidth]{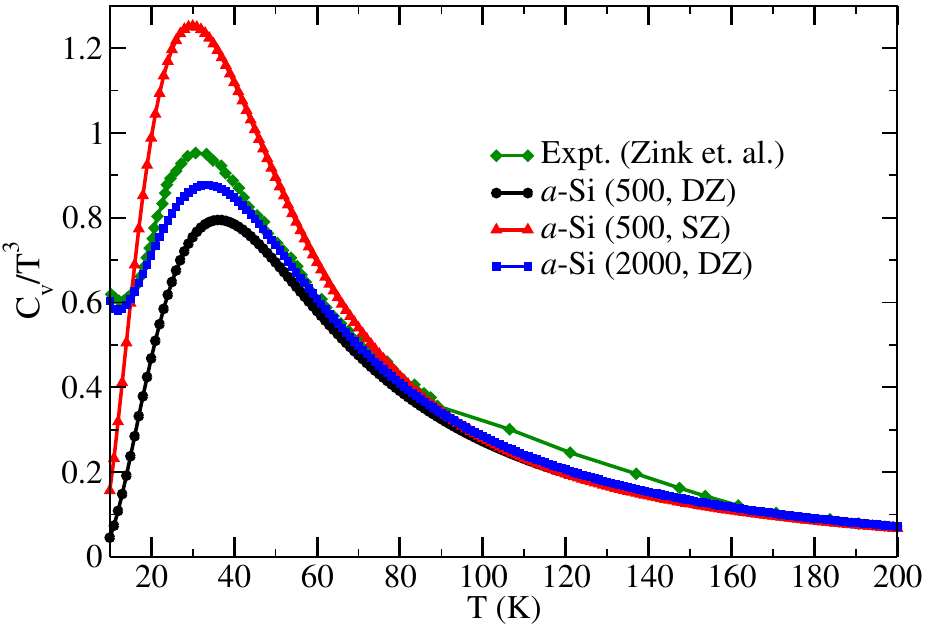}
\caption{
The dependence of the specific heat, $C_v$, on 
the SZ (red) and DZ (black) basis functions for 
500-atom models. Experimental data (green) and 
those from 2000-atom models (blue) using DZ 
basis functions are also included for comparison. 
The results are obtained by averaging over two 
independent configurations for each system size. 
}
\label{F17}
\end{figure}

We end this section with a brief discussion of the 
dependence of the MSD of atoms and the 
specific heat $C_v$ on the size of the basis 
functions and the nature of exchange-correlation 
(XC) approximations. Since vibrational excitations 
in solids typically involve energies of a few tens 
of meV, it is crucially important to calculate the 
elements of the dynamical matrix as accurately as 
possible by using an extended set of basis functions 
and a suitable XC functional appropriate for the 
system to be studied. To examine this issue, the MSDs 
of atoms in {\asi} were calculated using both 
single-zeta (SZ) and double-zeta (DZ) basis functions 
(from {\sc Siesta}), and the local density approximation 
(LDA) of the XC functional and its generalized-gradient 
counterpart (GGA).  The results of these calculations 
are shown in Fig.~\ref{F16}. It is evident from the 
plots that the use of SZ basis functions produces 
a somewhat larger value of {\u2} at all temperatures 
compared to its DZ counterparts for identical models. 
By contrast, the XC approximation has little or 
no effect on {\u2} values for a given basis set. 
Likewise, the computed values of $C_v$ are also found to 
be affected by the size of the basis functions used 
in the {\it ab initio} calculation of the vibrational 
frequency spectrum. This is apparent in Fig.~\ref{F17}, 
where the results for $C_v$ obtained from using SZ and 
DZ basis functions in {\sc Siesta} are presented for 
the 500-atom and 2000-atom models using the generalized 
gradient approximation. The results from 
Figs.~\ref{F16} and \ref{F17} are not surprising and 
they are a reflection of the fact that the accuracy of 
the total force acting on an atom -- and hence the 
elements of the dynamical matrix that are obtained 
from numerical derivatives of atomic forces with 
respect to atomic displacements -- depends on the 
size of the basis functions used in the calculations. 
In view of these findings, one may conclude that 
first-principles calculations of vibrational properties 
of amorphous silicon should be studied using an extended 
set of basis functions, whenever possible. 
Unfortunately, this requirement can considerably hinder 
our ability to carry out {\it ab initio} calculations 
of amorphous solids for large systems by constraining 
the size of the systems.

\section{Conclusions}

In this paper, we have studied the impact of the 
Debye-Waller factor on the structural and dynamical 
properties of~{\asi} using quantum-mechanical 
lattice-dynamical calculations in the harmonic 
approximation and from direct AIMD simulations. 
The quantitative effects of thermal vibrations on 
the intensity of the first two diffraction maxima, 
also known as the first sharp diffraction peak (FSDP) 
and the principal peak, are obtained by computing 
the MSD of atoms at 300~K and 600~K.   
The computed Debye-Waller factor at 300~K suggests 
that the intensities of the FSDP and the principal 
peak are reduced by a factor of 0.94 and 0.8, 
respectively, for CuK$_{\alpha}$ X-radiation.  
This observation is quite important and useful 
in comparing the room-temperature experimental 
structure-factor data of {\asi} with those from 
static model calculations, which do not account 
for thermal vibrations of atoms.

The squared displacement (SD) of atoms in the 
amorphous environment of silicon is found to 
vary from site to site and that it considerably 
depends on the degree of disorder in the local 
atomic environment, atomic connectivity,  and 
the coordination number of atoms. 
In particular, it has been observed that while 
the distribution of the local SDs from tetrahedral 
sites exhibits a nearly-gaussian behavior, the 
presence of dangling bonds, defects, inhomogeneities, 
and a few highly disordered sites (involving 
long Si-Si bonds) in amorphous networks can lead to a 
highly stretched asymmetric non-gaussian tail in 
the distribution. This non-gaussian behavior is distinctly different 
from that in elemental crystals, where thermal vibrations 
of atoms in an ordered atomic environment give 
rise to a gaussian distribution in the harmonic 
approximation.

The accuracy of the MSD values 
from lattice-dynamical calculations is examined by 
computing the MSD from direct AIMD calculations. 
The latter incorporates any anharmonicity that may be 
present in the atomic potential at temperature 
of up to 600~K. 
The average Debye temperature, estimated from the 
MSD of Si atoms in the Debye approximation, is 
found to be 541.5~K in the temperature range of 10--50~K, 
which is somewhat larger than the experimental 
value of $\Theta_d$ of 487--528~K obtained from 
specific-heat measurements at low temperatures. 
This observed 
deviation can be attributed to the absence of 
very low-frequency vibrational modes in small 
500-atom models, which underestimate the MSD 
of atoms and overestimate $\Theta_d$ in the 
Debye approximation via the inversion of 
Eq.~(\ref{EQ11}). However, a direct determination 
of $\Theta_d$ from specific-heat calculations 
using 2000-atom models of {\asi} is found to be 488~K. 
This value agrees very well with the experimental 
value of the Debye temperature of 487~K, reported 
by Zink et al.~\cite{Zink:2006}, but it is somewhat 
smaller than the value of 528~K obtained by Mertig et al.~\cite{Mertig:1984}
A comparison of $C_v$ from theory and experiments 
shows that the values 
obtained from the lattice-dynamical calculations 
match closely with those from experiments at 
temperatures above 40~K, but small deviations 
exist below 40~K. 

A review of experimental specific-heat data 
and theoretical values obtained from 
varying model sizes leads to the surmise that 
the absence of a small number of excess vibrational modes (relative to the 
Debye model) near 1 THz could be partly 
responsible for the deviation at low temperatures 
below 40~K. This statement appears to be vindicated 
in our study by investigating a four-fold increase 
of the system size from 500 atoms to 2000 atoms.  
However, a definitive answer to this 
surmise is outside the scope of the present study 
as it requires accurate calculations of the vibrational 
spectra using larger {\asi} models in order to examine 
the vibrational modes near the 1-THz region.  Such 
calculations are prohibitively difficult from an 
{\it ab initio} density-functional point of view due to the requirement 
of using an extended basis set to construct 
the force-constant matrix of {\asi} models involving several 
thousands of atoms in the self-consistent field 
approximation. The effects of anharmonicity on the 
atomic dynamics in {\asi} have been studied by computing the 
MSD from the equilibrium trajectories of Si atoms 
obtained from direct AIMD simulations in 
microcanonical and canonical ensembles.  
Comparisons of results from the direct AIMD and 
lattice-dynamical calculations show a small 
anharmonicity-induced increase of the 
MSD values at temperatures above 450~K.  
This suggests that the harmonic approximation works 
very well below 400~K, which is lower than the 
Debye temperature of 487--541~K for {\asi}. 

\vspace*{0.5 cm} 

\section{Acknowledgments}
The work was partially supported by the U.S. National
Science Foundation (NSF) under Grant No.\,DMR 1833035.
DD thanks the University of Southern Mississippi 
for financial support in the form a graduate assistantship. 

%\bibliography{dw.bib}

\begin{thebibliography}{44}
\bibitem [{DW()}]{DW}%
  \BibitemOpen
  \href@noop {} {}\bibinfo {note} {The Debye-Waller factor depends on the
  elemental species ($k$) of an atom and it is often written as the square of
  the temperature factor $T_k({\bf Q}) = \exp(-M_k) = \langle \exp[i{\mathbf Q}
  \cdot {\mathbf u}(k)] \rangle$, with $\langle\ldots\rangle$ indicating a time
  average. Assuming that the displacements are small (i.e., harmonic) and
  random, $2M$ is given by $\langle(\mathbf Q \cdot \mathbf u)^2\rangle$, where
  the vector ${\mathbf u}$ is parallel to ${\mathbf Q}$.}\BibitemShut {Stop}%
\bibitem [{DZ()}]{DZ}%
  \BibitemOpen
  \href@noop {} {}\bibinfo {note} {Although it is not necessary to use DZ basis
  functions for constructing {\asi} models, the specific heat and MSD of {\asi}
  from lattice-dynamical calculations are found to be dependent on the basis
  functions used to obtain the dynamical matrix in Eq.~(\ref{EQ2}). For
  numerical consistency, we therefore used the DZ basis functions throughout
  this work (see Figs.~\ref{F16} and \ref{F17}).}\BibitemShut {Stop}%
\bibitem [{sca()}]{scale}%
  \BibitemOpen
  \href@noop {} {}\bibinfo {note} {The argument can be readily presented for a
  one-dimensional ordered chain of lattice constant $a$ and spring constant
  $k$, where $\omega^2 (q\to0) = kq^2a^2/M$ in the nearest-neighbor interaction
  between atoms. A scale transformation $a' = s a $ leads to $q' = q/s$ and
  assuming $k$ remains unchanged for small $s$, we have $u_i^{'2} \approx
  \sum_n (k_BT)/(M\omega_n^{' 2}) \approx u_i^2$.}\BibitemShut {Stop}%
\bibitem [{U2()}]{U2}%
  \BibitemOpen
  \href@noop {} {}\bibinfo {note} {A numerical inversion of Eq.~(\ref{EQ11})
  shows that the value of $x_d$ decreases rapidly as $u_d^2$ increases. This
  follows from the structure of the integral in (\ref{EQ11}), and it is
  independent of the approximation, $u_d^2 \approx \langle u^2 \rangle$, used
  to obtain $\Theta_d$ from the MSDs.}\BibitemShut {Stop}%
\end{thebibliography}

\end{document}